\documentclass[11pt,twocolumn]{aastex}
\usepackage{natbib}
\usepackage{graphicx}
\usepackage{wrapfig}
\usepackage[caption=false]{subfig} 
 \pdfoutput=1
\usepackage{setspace}
\usepackage{psfig}
\usepackage[utf8x]{inputenc}
\usepackage{xcolor}

\shortauthors{Karnath et al.}

\begin{document}

\title{
 Detection of Irregular, Sub-mm Opaque Structures in the Orion Molecular Clouds: Protostars within 10000 years of formation?}

\author{N. Karnath \altaffilmark{1,11}}
\author{S.~T. Megeath \altaffilmark{1}}
\author{J.~J. Tobin \altaffilmark{2}}
\author{A. Stutz \altaffilmark{3,4}}
\author{Z.-Y. Li \altaffilmark{5}}
\author{P. Sheehan \altaffilmark{2}}
\author{N. Reynolds \altaffilmark{6}}
\author{S. Sadavoy \altaffilmark{7}}
\author{I.~W. Stephens \altaffilmark{8}}
\author{M. Osorio \altaffilmark{9}}
\author{G. Anglada \altaffilmark{9}}
\author{A.~K. D\'{i}az-Rodr\'{i}guez \altaffilmark{9}}
\author{E. Cox \altaffilmark{10}}

\altaffiltext{1}{Ritter Astrophysical Research Center, University of Toledo, 2801 W. Bancroft Street, Toledo, OH 43606, USA}
\altaffiltext{2}{National Radio Astronomy Observatory, 520 Edgemont Road Charlottesville, VA 22903-2475, USA}
\altaffiltext{3}{Departmento de Astronom\'{i}a, Facultad de Ciencias F\'{i}sicas y Matem\'{a}ticas, Universidad de Concepci\'{o}n, Concepci\'{o}n, Chile}
\altaffiltext{4}{Max-Planck-Institute for Astronomy, K\"onigstuhl 17, 69117 Heidelberg, Germany}
\altaffiltext{5}{Department of Astronomy, University of Virginia, Charlottesville, VA 22903}
\altaffiltext{6}{Homer L. Dodge Department of Physics and Astronomy, University of Oklahoma, 440 W. Brooks Street, Norman, OK 73019, USA}
\altaffiltext{7}{Department of Physics, Engineering Physics, and Astronomy, Queen's University, Kingston, ON K7L 3N6 Canada}
\altaffiltext{8}{Harvard-Smithsonian Center for Astrophysics, 60 Garden Street, Cambridge, MA, USA}
\altaffiltext{9}{Instituto de Astrof\'{i}sica de Andaluc\'{i}a, CSIC, Glorieta de la Astronom\'{i}a, s/n, 18008 Granada, Spain}
\altaffiltext{10}{Center for Interdisciplinary Exploration and Research in Astrophysics (CIERA), Department of Physics \& Astronomy, Northwestern University, 2145 N. Sheridan Rd., Evanston, IL 60208, USA}
\altaffiltext{11}{Current Address: SOFIA Science Center, USRA, NASA Ames Research Center, Moffett Field, CA 94035, USA; nkarnath@sofia.usra.edu}

\begin{abstract}

We report ALMA and VLA continuum observations that potentially identify the four youngest protostars in the Orion Molecular Clouds taken as part of the Orion VANDAM program.  These are distinguished by bright, extended, irregular emission at 0.87~mm and 8~mm and are optically thick at 0.87~mm. These structures are distinct from the disk or point-like morphologies seen toward the other Orion protostars. The 0.87~mm emission implies temperatures of 41-170~K, requiring internal heating. The bright 8~mm emission implies masses of 0.5 to 1.2 M$_{\odot}$ assuming standard dust opacity models.  One source has a Class 0 companion, while another exhibits substructure indicating a companion-candidate. Three compact outflows are detected, two of which may be driven by companions, with dynamical times of $\sim$300 to $\sim$1400 years. The slowest outflow may be driven by a first hydrostatic core. These protostars appear to trace an early phase when the centers of collapsing fragments become optically thick to their own radiation and compression raises the gas temperature. This phase is thought to accompany the formation of hydrostatic cores.  A key question is whether these structures are evolving on free fall times of $\sim\,$100 years, or whether they are evolving on Kelvin-Helmholtz times of several thousand years. The number of these sources imply a lifetime of $\sim$6000 years, in closer agreement with the Kelvin-Helmholtz time.  In this case, rotational and/or magnetic support could be slowing the collapse.

\end{abstract}


\section{Introduction}
\label{sec:intro}

The transition from a dense collapsing clump of gas and dust in a fragmenting cloud to an accreting protostar is perhaps the most challenging phase of a star's birth to observe. This transition occurs within the dense centers of these collapsing objects, where increases in temperature and pressure slow collapse and ultimately lead to the formation of opaque hydrostatically supported cores (hereafter: HSCs\footnote{We refer to the dense, isothermal regions of clouds  undergoing collapse as fragments and the hydrostatically supported objects in the centers of the fragments as cores.  In the literature, the phrase ``molecular core" is often used to describe what we refer to as fragments, e.g., \citet{bergin2007}.}). Theoretical studies have linked this transition to high optical depth. Early studies on opacity limited star formation produced models that predicted hierarchical Jeans fragmentation of clouds would cease when the fragments became optically thick to their own radiation \citep{low1976,rees1976,silk1977}. Numerical simulations following the collapse of spherical, constant density fragments showed that the formation of an optically thick region precedes the formation of HSCs \citep[e.g., ][]{larson1969}.  In these studies \citep[e.g., ][]{masunaga1998}, the initial, isothermal collapse was characterized by a contracting, constant density, central region with a diameter approximately equal to the Jeans length surrounded by a $\rho \propto r^{-2}$ envelope. \citet{masunaga1998} argued that a HSC forms when heating by compression in the limit of constant density exceeds cooling by radiation. They found that this transition occurs at approximately $10^{-13}$~g~cm$^{-3}$, the density at which the core becomes optically thick to its own radiation. 

Numerical simulations of this transition have been ongoing for decades, including \citet{larson1969, masunaga1998, masunaga2000, commercon2011, vaytet2013, bate2014, bhandare2018} and others. \citet{larson1969} was the first numerical study to show that an increase in temperature and pressure led to the formation of a very low mass, first hydrostatic core (hereafter:~FHSC), comprised of molecular hydrogen. As the accreting FHSC grows in mass and the central temperature reaches $\sim$2000~K, the dissociation of the H$_{2}$ molecules causes the FHSC to collapse and form the second hydrostatic core. The appearance of the second HSC is expected to mark the onset of the Class 0 protostar phase, and the central, hydrostatically supported protostar  will continue to grow in mass as it accretes from the infalling envelope and disk. The work of \citet{larson1969} assumed initial conditions of spherical symmetry, constant density, and no rotation, turbulence, or magnetic fields. Since Larson's work, many contributions have shown that magnetic fields \citep{basu1994, tomisaka2002, commercon2011} and rotation \citep{tsuribe1999a,tsuribe1999b} can influence collapse and lead to the formation of disks or disk-like structures and outflows during the FHSC phase, i.e., before the formation of the protostar \citep[e. g.][]{saigo2000, matsumoto2003, bate2014}. 


Observations of this transition are challenging due to the high optical depths at the centers of the fragment, i.e., opaque regions where the HSCs form, the small spatial extent of the nascent HSCs, and their rapid evolution. With the sensitivity and resolution of sub-mm/mm interferometers such as ALMA and the VLA, however, we can now detect and resolve the dense, opaque regions of collapsing fragments. The observations presented in this paper improve our understanding of star formation in a regime that has been previously restricted to theoretical investigations. A couple opaque core candidates have been found in the literature thus far, for example, \citet{cox2015,hernandezgomez2019}; these demonstrate the need for long wavelength observations with ALMA and VLA to detect such deeply embedded objects.

Given the rapid evolution of this transitional phase, surveys of hundreds of protostars are required to detect even a few examples. The VLA/ALMA Nascent Disk and Multiplicity (VANDAM) survey in Orion (Tobin et al. accepted) observed 328 $Herschel$ Orion Protostellar Survey \citep[HOPS;][]{fischer2013, furlan2016} protostars with ALMA at 0.87~mm and 148 protostars with VLA at 8~mm and 1~cm with $\sim$40~AU spatial resolution. These include the 19 extremely red (70~$\mu$m) protostars, or ``PACS Bright Red Sources" \citep[hereafter: PBRs,][]{stutz2013}. Using imaging from the $Spitzer$ Space Telescope and the $Herschel$ Space Observatory, \citet{stutz2013} identified this sample of 18 protostars; one additional PBR has been identified since \citep{tobin2015}. They are distinguished by their weak mid-IR emission and red 70 $\mu$m to 24 $\mu$m colors; eleven of the PBRs were not identified as protostars by $Spitzer$ due to their faint 24~$\mu$m fluxes, and eight were not detected by $Spitzer$ at $\lambda \le 24$~$\mu$m.  Subsequent CARMA observations showed the PBRs had high L$_{2.9~mm}$/L$_{bol}$ ratios compared to other Class 0 protostars \citep{tobin2015}. Using radiative transfer models to demonstrate the PBRs were internally heated, \citet{stutz2013} interpreted the PBRs as very young Class 0 protostars. The red colors and rarity of PBRs suggest they represent an early phase of protostellar evolution with a duration of $\sim$10,000 years \citep{stutz2013}.


We present here the detection of four previously identified PBRs (HOPS~400-B, HOPS~402, HOPS~403, and HOPS~404) that show extended, irregular morphologies in VLA (8~mm) and ALMA (0.87~mm) imaging unlike the vast majority of the VANDAM protostar sample. These irregular PBRs are of particular interest because of their large and asymmetric structures. Two additional protostars are discussed to a smaller extent, a Class 0 protostellar companion to an extended PBR (HOPS~400-A) and a more compact PBR in spatial proximity to an extended PBR (HOPS~401). All of the PBRs presented here are found in Orion B and we adopt a common distance of 400~pc for all the protostars in this study \citep{kounkel2018}. We find that the inner, $\sim$160~AU regions resolved by ALMA and the VLA are optically thick at 0.87 mm, indicating that they may be the theoretically predicted, opaque regions at the earliest stages of HSC formation. Using the masses, luminosities, and number of the optically thick regions, as well as detections of outflows in $^{12}$CO, we assess the timescales of these protostars and compare their properties to those anticipated by simulations.  We find that these observations challenge current models that predict small (few~AU) optically thick regions and thermally supported cores. We discuss possible solutions for resolving these challenges. 

The paper is organized as follows: in \S \ref{sec:observations}, we describe the observations and reduction of the data presented, \S \ref{sec:results} contains the observational results, \S \ref{sec:discussion} presents the discussion and possible scenarios for varying collapse timescales and support mechanisms, the summary is in \S \ref{sec:summary}, and Appendix A contains the details of each source.

\section{Observations and Data Reduction}
\label{sec:observations}
\subsection{ALMA 0.87~mm Observations and Reduction}

The Cycle 4 ALMA 0.87~mm observations were conducted on 2016 September 4 and 5, and 2017 July 19 for which 34, 39, and 42 antennas were operating, respectively, with baselines between 15.1~m to 3697~m. The total time spent on each target was $\sim$0.9 minutes.

The correlator was configured with two basebands set to 1.875 GHz bandwidth, centered at 333 GHz and 344 GHz. One of the remaining basebands was centered on $^{12}$CO ($J=3\rightarrow2$) at 345.79599 GHz with a total bandwidth of 937.5~MHz and 0.489 km s$^{-1}$ channels. The last remaining baseband was centered on $^{13}$CO ($J=3\rightarrow2$) at 330.58797~GHz, with a bandwidth of 234.375~MHz and 0.128 km s$^{-1}$ channels. The line free-regions of the basebands were used for additional continuum bandwidth. The total aggregate continuum bandwidth was $\sim$4.75~GHz.

The data were manually reduced and imaged using the Common Astronomy Software Application \citep[CASA][]{mcmullin2007} by the Dutch Allergro ARC Node. We used CASA 4.7.2 for all self-calibration and imaging. Following the standard calibration, up to three rounds of phase-only self-calibration were performed on the continuum data to increase the S/N. The self-calibration solutions were also applied to the spectral line data. Final continuum and spectral line data cubes were then produced.

The continuum images were deconvolved using Briggs weighting with a robust parameter of 0.5, yielding a synthesized beam of 0\farcs11$\times$0\farcs10 (44~AU~$\times$~40~AU). The continuum image also only uses uv-points at baselines $>$25~k$\lambda$ to mitigate striping resulting from large-scale emission that is not properly recovered. The spectral line data were deconvolved using Natural weighting for baselines $>$50 k$\lambda$ with outer taper of 500~k$\lambda$ applied to increase the sensitivity to extended structure, yielding synthesized beams of 0\farcs25$\times$0\farcs24. The continuum subtraction was done using line-free regions of the spectral windows to estimate the continuum level using the $uvcontsub$ task. The resulting RMS noise of the continuum images, $^{12}$CO datacubes, and $^{13}$CO datacubes are $\sim$0.31~mJy~beam$^{-1}$, 17.7 mJy~beam$^{-1}$ (1 km s$^{-1}$ channels), and 33.3 mJy~beam$^{-1}$ (0.44 km s$^{-1}$ channels), respectively. The flux density of the objects and surrounding extended structure are measured both using Gaussian fitting and a measurement within a polygon region, these yield measured flux densities that are within the uncertainty of one another. For consistency with Tobin et al. (accepted), we use the Gaussian fitting results in our tables and analysis. The peak and integrated flux densities with uncertainties, and positions are given in Table \ref{tbl:pbrsobservables}. The absolute flux calibration accuracy is expected to be $\sim$10\%, and comparisons of the observed flux densities for the science targets for different executions are consistent with this level of accuracy. The uncertainties in Table \ref{tbl:pbrsobservables} are the statistical uncertainties and do not include the calibration error.

\subsection{VLA Observations and Reduction}

The observations with the VLA were conducted in the A-configuration on October 21, 23, 29, and 31 and December 19, 2016. Each target typically was observed individually in a single execution with $\sim$1~hr on-target. The observations used the Ka-band receivers and the correlator was used in the wide bandwidth mode (3-bit samplers) with one 4~GHz baseband centered at 36.9~GHz (8.1~mm) and the second 4~GHz baseband was centered at 29~GHz (1.03~cm). We chose to use the 8.1~mm data over the combination of 8.1~mm and 1~cm data to maximize the dust emission over a smaller bandwidth and the 8.1~mm data has higher flux densities than 1~cm, which is better for studying the dust emission. 

The data were reduced using the scripted version of the VLA pipeline in CASA 4.4.0. The continuum was imaged using the \textit{clean} task in CASA 4.5.1 using Natural weighting, multi-frequency synthesis with \textit{nterms=1} across both basebands. The synthesized beams are 0\farcs08$\times$0\farcs07 (32~AU $\times$ 28~AU). 

The flux density of the objects and surrounding extended structure are measured using both Gaussian fitting and integrating the intensity over a polygonal region just as found for the ALMA data, these two approaches yield results that are within the uncertainties of one another. Using the Gaussian value, the peak and integrated flux densities with statistical uncertainties, and positions are given in Table \ref{tbl:pbrsobservables}. The absolute flux calibration accuracy of the VLA is expected to be $\sim$10\%. The uncertainties in Table \ref{tbl:pbrsobservables} are the statistical uncertainties and do not include the calibration error.

\begin{deluxetable*}{cccccccc}
\tablewidth{0pt}
\tablecaption{Observed Properties of Sources
 \label{tbl:pbrsobservables}}
\tablehead{
\colhead{$ $} & \colhead{$ $} & \multicolumn{2}{c}{Centroid Positions} & \multicolumn{2}{c}{VLA (8.1~mm)} & \multicolumn{2}{c}{ALMA (0.87~mm)}  \\
\colhead{HOPS ID} & \colhead{Type of } & \colhead{RA$^{a}$} & \colhead{Dec$^{a}$} & \colhead{Flux Density} & \colhead{Intensity Peak} &\colhead{Flux Density} & \colhead{Intensity Peak}  \\
\colhead{$ $} & \colhead{PBR} & \colhead{(J2000)} & \colhead{(J2000)} & \colhead{(mJy)} & \colhead{(mJy/beam)}  & \colhead{(mJy)} & \colhead{(mJy/beam)}}
\startdata
400-A & regular & 05:42:45.25 & -01:16:13.64 & 0.802$\pm$0.020 & 0.524$\pm$0.008 & 147.7$\pm$7.3 & 67.9$\pm$2.0 \\
400-B & irregular & 05:42:45.26 & -01:16:14.14 & 2.598$\pm$0.088 & 0.254$\pm$0.008 & 657.1$\pm$19.3 & 49.4$\pm$2.0 \\
401 & regular & 05:46:07.65 & -00:12:20.73 & 0.328$\pm$0.028 & 0.146$\pm$0.007 & 128.9$\pm$2.9 & 40.1$\pm$0.6 \\
402 & irregular & 05:46:09.97 & -00:12:16.85 & 2.117$\pm$0.093 & 0.149$\pm$0.008 & 419.4$\pm$9.8 & 35.5$\pm$1.1 \\
403 & irregular & 05:46:27.75 & -00:00:53.81 & 3.803$\pm$0.105 & 0.295$\pm$0.008 & 584.1$\pm$12.8 & 41.6$\pm$1.3 \\
404 & irregular & 05:48:07.76 & 00:33:50.79 & 3.333$\pm$0.059 & 0.697$\pm$0.011 & 306.4$\pm$6.2 & 99.5$\pm$1.5 \\
\enddata
\tablenotetext{a}{The positions are based on the VLA maps.}
\tablenotetext{b}{The reported uncertainties are statistical uncertainties and do not include the calibration error of $\sim$10\% in both VLA and ALMA.}
\end{deluxetable*}

\section{Results}
\label{sec:results}

Although all 19 currently known PBRs were observed by the VLA and ALMA, we focus primarily on the four with bright emission at 8~mm and extended, irregular-shaped morphologies in both wavelengths: HOPS 400-B, 402, 403, and 404 \citep{furlan2016}. These four PBRs are unique amongst not only the PBRs sample but the entire VANDAM sample: their large and asymmetric structures set these sources apart. Two nearby protostars are also included in this analysis: HOPS~400-A is a PBR and a Class 0 companion to the irregular PBR, HOPS~400-B, and HOPS~401, which is a very red PBR. HOPS~401 appears to be a more compact, potentially irregular source and is found within 15000~AU of the irregular PBR, HOPS~402.  We tabulate and display the properties of HOPS 400-A and HOPS~401 in this section, but we defer discussion of those sources to \S \ref{sec:discussion} and \S \ref{sec:appendix}.

\begin{figure*}
\centering
\plotone{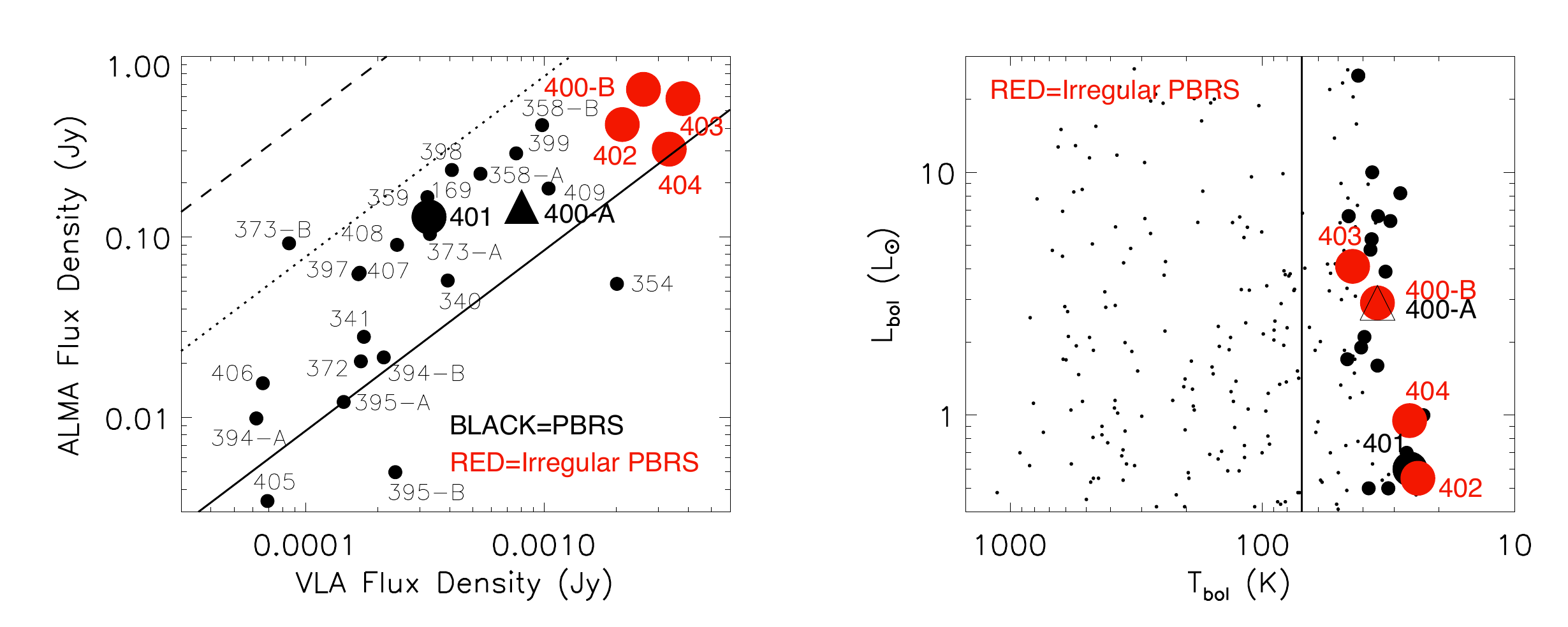}
\caption{\textbf{Left:} ALMA flux density vs. VLA flux density for the PBRs in the VANDAM Orion survey. The large red points are the four irregular PBRs; HOPS~401 and HOPS 400-A are the large black point and triangle, respectively. The smaller black points are the remaining PBRs sample. The overlaid lines are spectral indices between 8~mm and 0.87~mm of 2 (solid), 3 (dotted), and 3.8 (dashed). The four opaque irregular PBRs are bright at both wavelengths making them stick out from the rest of the sample. \textbf{Right:} L$_{bol}$ vs. T$_{bol}$ for the irregular PBRs (red) plus HOPS~401 and HOPS~400-A as the large black point and triangle, respectively. The medium-sized black points are the remaining PBRs sample and the small black points are the entire Orion VANDAM sample (Class 0 to Flat Spectrum protostars). The vertical line at 70~K delimits the Class 0 sources from the Class I sources \citep[e.g.][]{furlan2016}. The irregular PBRs fit into the rest of the PBRs sample and do not stick out in L$_{bol}$ vs. T$_{bol}$ space, aside from having among the lowest T$_{bol}$.}
\label{fig:fluxes}
\end{figure*}

\subsection{Flux Densities and Morphologies}

Figure \ref{fig:fluxes} shows the 0.87 and 8~mm flux densities, via Gaussian fits for consistency, for the entire PBRs sample of \citet{stutz2013} and \citet{tobin2015}. The four irregular PBRs that are the focus of this paper, HOPS~400-B, 402, 403 and 404, are located on the upper right. Their 0.87~mm flux densities are among the highest of the sample and their 8 mm flux densities exceed those of all the PBRs. The four irregulars in our sample have a spectral index of 2 to 3 between 0.87~mm and 8~mm, indicating thermal dust emission. However, HOPS-354 has a spectral index lower than 2, suggesting that there is significant free-free emission in this source. The four irregular PBRs in the upper right of Figure \ref{fig:fluxes} are also above the median 2.9~mm flux density of the PBRs \citep{tobin2015}. This is consistent with the ALMA and VLA flux densities being dominated by strong thermal dust emission and supports the interpretation that the irregular PBRs have the most massive, densest inner envelopes of the 19 PBRs.  

The bolometric luminosity and temperature (BLT) diagram for the PBRs is also presented in Fig.~\ref{fig:fluxes}. The L$_{bol}$ and T$_{bol}$ values of the four irregular protostars are consistent with the other PBRs \citep[see also][Fig. 14]{stutz2013}. Two of the irregular protostars, HOPS~402 and HOPS~404, and the protostar HOPS~401 have relatively low values for T$_{bol}$ and L$_{bol}$ compared to all HOPS Class~0 protostars \citep{fischer2017}. As described later, the presence of more evolved companions that are unresolved in the IR data may explain the higher values of T$_{bol}$ and L$_{bol}$ for HOPS~400 and perhaps also for HOPS~403.  

The continuum images of the PBRs sample studied in this paper are shown in Figure~\ref{fig:hopsimages}. See Appendix A for more details of each source. HOPS 400 is a resolved binary and we refer to the northern, compact Class 0 PBR as 400-A and the southern, extended, irregular companion as 400-B. HOPS~402 is asymmetric in both 0.87~mm and 8~mm images with no point-like peak resolved at 8~mm. HOPS~403 shows a point-like peak at 8~mm that is coincident with the more elongated, lower contrast 0.87~mm peak, and a second weaker peak at 8~mm. We classify HOPS~404 as an irregular since it shows a square shape with an offset peak at 8~mm, although it has more circular morphology at 0.87~mm. Overall, the four irregular PBRs are more extended and show highly structured emission in the 0.87~mm images than the rest of the VANDAM sample (Tobin et al. accepted). The extended, irregular emission at 0.87~mm and 8~mm  separates them from the remainder of the PBRs and other Class 0 protostars (Tobin et al. accepted).

\subsection{Brightness Temperature Profiles}
\label{sec:temps}

Assuming the emission lies within the Rayleigh-Jeans limit, Figure \ref{fig:temps} plots brightness temperature profile cuts at 0.87~mm and 8~mm along R.A. and are centered on the 8~mm peaks. Given a constant dust temperature, the ratio of the brightness temperatures is given by the equation

\begin{equation}
\label{eqn:temp}
		\frac{T_{b}(8~{\rm mm})}{T_{b}(0.87~{\rm mm})} =  \frac{1-e^{-\tau_{0.87}\left(\frac{0.87~{\rm mm}}{8~{\rm mm}}\right)^{\beta}}}{1-e^{-\tau_{0.87}}},
\end{equation}

\noindent
where $\tau_{870}$ is the optical depth at 0.87~mm and $\beta$ is the dust opacity index. For  $\beta$ ranging from 1.8 to 1, which is typical in protostellar envelopes \citep{ossenkopf1994,ormel2011}, the expected brightness temperature ratios should be between $\sim 0.01$ to $\sim 0.1$. When the ratio of the 8~mm to 0.87~mm brightness temperature exceeds 0.16 assuming $\beta=1$ (or exceeds 0.03 with $\beta=1.8$), the 0.87~mm emission is becoming optically thick ($\tau >$ 1). In Figure \ref{fig:temps}, vertical lines indicate where $\tau_{0.87~mm}$ exceeds 1 and the 0.87~mm emission is becoming optically thick for $\beta=1$. Since the ratio of 8~mm to 0.87~mm brightness temperature exceeds 0.16 over much of the detected emission, we conclude that the 0.87~mm emission is mostly optically thick toward all four of the irregular PBRs. The exact limits of the optically thick region depends on the adopted dust law.

The high optical depth at 0.87~mm is also apparent in the shapes of the cuts, which exhibit more small-scale structure at 8~mm than 0.87~mm. Peak brightness temperatures of the 8~mm emission exceed those of the 0.87~mm emission for three of the four irregular PBRs, with HOPS~402 as the exception. Toward the peak there is a combination of resolved and unresolved emission. The unresolved emission likely contains  even higher peak brightness temperatures than the maximum values found in the maps.  The high brightness temperatures at 8mm are clear evidence of temperature gradients in HOPS~400-B, HOPS~403, and HOPS~404. A similar temperature gradient may be present in HOPS~402, but the 8~mm temperature never exceeds the 0.87~mm temperature. The presence of temperature gradients (temperature decreasing with radius) demonstrate that there is internal heating.   

\begin{figure*}
\centering
\plotone{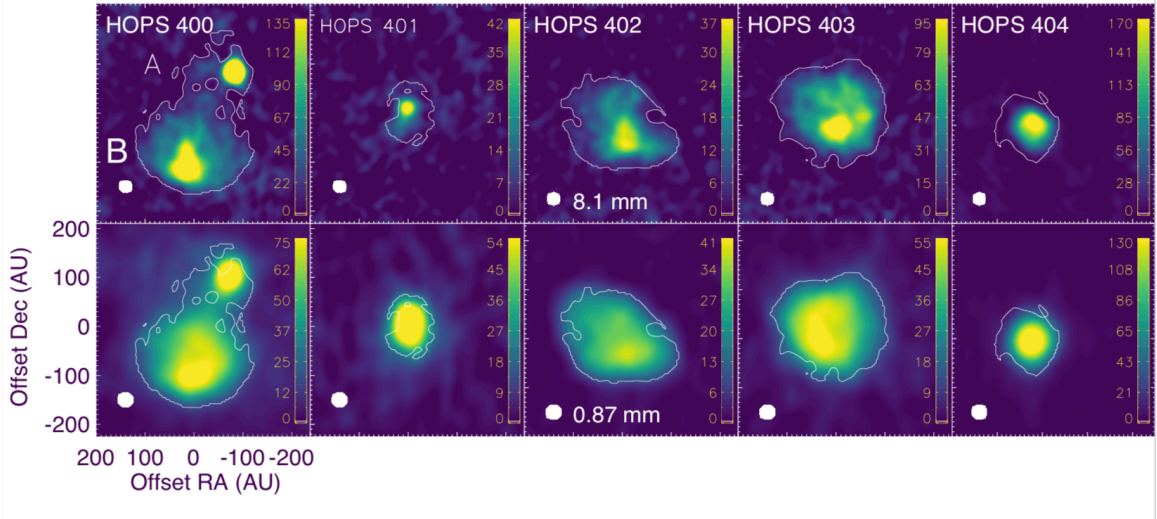}
\caption{Continuum images with VLA (8.1~mm, top) and ALMA (at 0.87~mm, bottom) of HOPS 400-404. HOPS~401 is discussed in Appendix A. The colorbars represent the brightness temperature scales in each individual panel. The contours represent where $\tau$=1 (see Eqn \ref{eqn:temp}) for the 0.87~mm continuum emission assuming $\beta$=1. The coordinates are oriented with RA-Dec (J2000) at a distance of 400~pc. The synthesized beams are overlaid, FWHM=42~AU in ALMA and 30~AU in VLA.}
\label{fig:hopsimages}
\end{figure*}

\subsection{Temperatures, Radii, and Masses}
\label{sec:mass}

To characterize the protostars, we estimate their average dust temperatures ($T_{d}$) at the 0.87~mm photosphere, their radii, and  masses. Given the irregular shapes, we determine a half max radius using r$_{hm}$=$\sqrt{A/\pi}$, where $A$ is the area within the half maximum contour. To account for beam convolution, we subtract the beam radius at 400~pc in quadrature to calculate the deconvolved radius (r$_{decon}$). The deconvolved radii range from 29 to 84~AU (Table \ref{tbl:pbrsproperties}). 

To estimate the dust temperature, we use the average 0.87~mm brightness temperature for pixels within the half maximum of the 0.87 mm emission and assume the 0.87~mm emission is optically thick everywhere within this contour. The resulting dust temperatures, given in Table~\ref{tbl:pbrsproperties}, range from 33 to 95~K. These are higher than the bolometric temperatures estimated from the SEDs \citep[24 - 44~K,][]{stutz2013,furlan2016}, and are higher than the typical dust temperature found in the dense gas of the Orion~B cloud \citep[16~K,][]{schneider2013}. This is additional evidence that these sources are internally heated. If there was no internal heating, we expect dust temperatures that would be comparable to the typical dust temperature found in \citet{schneider2013}. Additionally, the SEDs in \citet{stutz2013,furlan2016} are modified blackbody models fit to single dish data with 7-19$\arcsec$ angular resolution; these contain contributions from the colder, extended envelopes surrounding the structures detected by ALMA/VLA.  


We calculate the dust mass for the irregular PBRs by assuming that the 8~mm emission is optically thin. An additional assumption is that the 8~mm emission traces the dust, with a negligible free-free contribution. We adopt the average dust temperature within r$_{decon}$ of the 0.87~mm emission as the overall dust temperature. The dust temperature is combined with the integrated flux density at 8~mm to calculate  
\\
\noindent	
	\begin{equation}
		M_{dust} = \frac{F_{\nu}d^{2}}{\kappa_{\nu}B_{\nu}(T_{d})}.
	\end{equation}
	\noindent
B$_{\nu}$(T$_{d}$) is the blackbody intensity at the average dust temperature T$_{d}$ from the 0.87~mm data. 

As a fiducial dust opacity, we adopt $\kappa_{8.1}$ = 0.144 cm$^{2}$ g$^{-1}$ used by \citet{segura2016, tychoniec2018} and Tobin et al. (accepted). This value comes from the 1.3~mm opacity of 0.899 cm$^{2}$ g$^{-1}$ \citep{ossenkopf1994} extrapolated  to 8.1~mm using a $\beta$=1 as the dust opacity spectral index. The dust masses are multiplied by 100 to estimate the total gas mass presented in Table~\ref{tbl:pbrsproperties}. The masses derived here do not include the mass in a central HSC, if present, which will be much smaller than the beam and optically thick at 8~mm. The masses may be affected by optical depth at 8~mm, which will result in an underestimate of the mass, and by higher temperatures in the opaque regions, which will result in an over estimate. 

The largest uncertainty in the masses comes from the range of possible dust opacities. We give a plausible range of $\kappa$ values in Table \ref{tbl:kappa} using different dust laws in the literature. Our fiducial dust law falls above \citet{ossenkopf1994} dust law with an initial MRN size distribution, thin ice mantles and a gas density of 10$^{6}$ cm$^{-3}$, this law gives a kappa value of 0.05 cm$^{2}$ g$^{-1}$ when extrapolated to 8~mm and increases the resulting masses of the optically thick regions by a factor of $\sim$3. On the opposite end, extrapolating the \citet{ossenkopf1994} law for an initial MRN distribution, no ice, and a gas density of 10$^{8}$ cm$^{-3}$ yields $\kappa$=0.76 cm$^{2}$ g$^{-1}$, which lowers the resulting mass by a factor of $\sim$5. 

Clearly the adopted opacity law will change the resulting mass and influence the interpretation of the irregular PBRs. We conclude that the uncertainty in mass from the dust law is over one order of magnitude. Further uncertainties in the mass may come from variations in the gas to dust ratio, we adopt 100:1, and adopted dust temperature \citep[e.g.,][]{bate2017}. The footnotes of Table~\ref{tbl:pbrsproperties} describe how the calculated values vary with the adopted opacity, and we discuss the implication of this range in the next section. We note that the masses calculated with the optically thick 0.87~mm data tabulated in Tobin et al. (accepted) underestimate the more accurate masses calculated with the 8~mm using the same dust law ($\kappa$=0.144 cm$^{2}$ g$^{-1}$).

\begin{deluxetable*}{ccccccccccc}
\tablewidth{0pt}
\tablecaption{Physical Properties of Sources
 \label{tbl:pbrsproperties}}
\tablehead{
\colhead{HOPS ID} & \colhead{T$_{d}$} & \colhead{M$^{a,b}$} & \colhead{r$_{decon}$} & \colhead{L$_{bol}^{c}$} & \colhead{L$_{tot}^{c}$} & \colhead{L$_{op}$} & \colhead{T$_{bol}^{c}$} & \colhead{log($\rho$)} & \colhead{log(n)} &  \colhead{T/U$^{a,b,d}$} \\
\colhead{$ $} & \colhead{(K)} & \colhead{(M$_{\odot}$)} & \colhead{(AU)} & \colhead{(L$_{\odot}$)} & \colhead{(L$_{\odot}$)} & \colhead{(L$_{\odot}$)} & \colhead{(K)} & \colhead{(g cm$^{-3}$)} & \colhead{(cm$^{-3}$)} & \colhead{$ $}}
\startdata
400-B & 48 & 0.70 & 54.9 & 2.94 & 5.2 & 0.65 & 35 & -12.22 & 11.11 & 0.018 \\ 
401 & 39 & 0.11 & 28.7 & 0.61 & 0.75 & 0.08 & 26 & -12.18 & 10.68 & 0.047 \\ 
402 & 33 & 0.83 & 82.9 & 0.55 & 0.6 & 0.33 & 24 & -12.68 & 10.65 & 0.015 \\ 
403 & 41 & 1.20 & 84.1 & 4.1 & 5.3 & 0.82 & 44 & -12.54 & 11.79 & 0.013 \\ 
404 & 95 & 0.45 & 32.9 & 0.95 & 1.5 & 3.6 & 26 & -11.75 & 11.59 & 0.032 \\ 
\enddata
\tablenotetext{a}{Adopting $\kappa =0.144$~cm$^{2}$~g$^{-1}$ (see \S \ref{sec:mass})}
\tablenotetext{b}{Value scales with opacity by equation $M = (0.144~{\rm cm}^{2}~{\rm g}^{-1}/\kappa_{8mm})\times$ M(tabulated) (see \S \ref{sec:mass} and \ref{sec:energy})}
\tablenotetext{c}{Furlan et al. (2016)}
\tablenotetext{d}{Scales by T/U = $\kappa_{8mm}$ / 0.144 $cm^{2}$ $g^{-1}$ $\times$ T/U (tabulated) (see \S \ref{sec:energy})}
\end{deluxetable*}

\begin{figure*}
\centering
\plotone{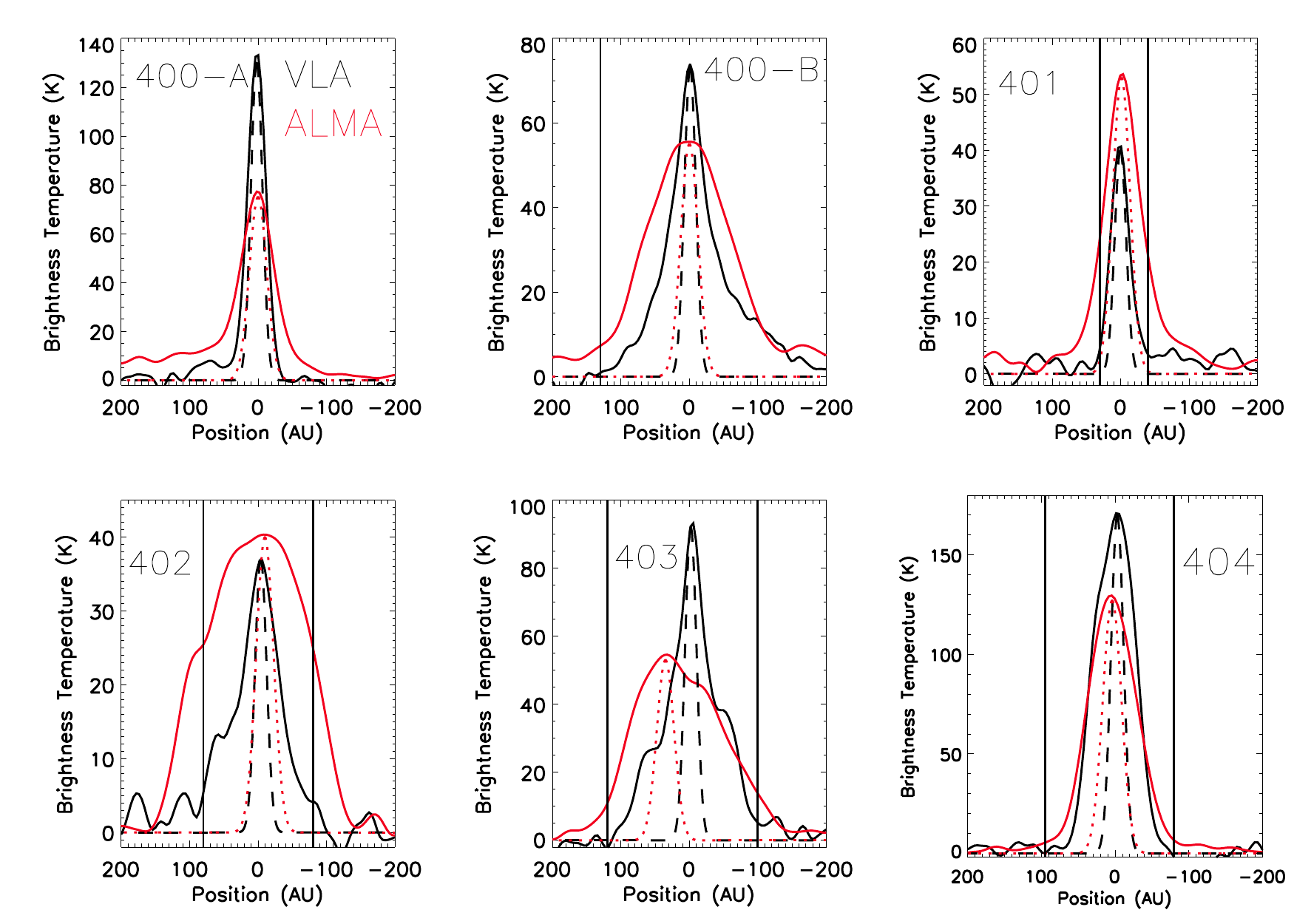}
\caption{Brightness temperature profiles of the ALMA continuum images (red, 0.87~mm) and of the VLA continuum images (black, 8~mm) for HOPS 400-404. The cuts are made across the image in right ascension at the peak of the 8~mm emission with postivie RA position corresponding to increasing RA (east on right to west on left). The red (dotted) and black (dashed) lines are the normalized beam sizes for ALMA and VLA, respectively. The solid vertical lines indicate where the 0.87~mm emission is becoming optically thick ($\tau > 1$) assuming $\beta=1$. Note for HOPS 400-B the optically thick 0.87~mm emission occurs at -220~AU and therefore is not visible in the plot.}
\label{fig:temps}
\end{figure*}

\subsection{Luminosities and Timescales} 
\label{sec:timescales}

The $L_{bol}$ (integrated SED including both internal and external heating) and $L_{tot}$ (luminosity from model fits including corrections for foreground extinction and beaming by outflow cavities) derived luminosities for the protostars given by \citet{furlan2016} range from 0.6 - 2.9 L$_{\odot}$ and from 0.6 - 5.3 L$_{\odot}$, respectively (Table \ref{tbl:pbrsproperties}). Note that L$_{tot}$ makes assumptions about the structure of the protostellar envelope as well as the lack of external heating, which may not be appropriate for the four irregular PBRs.  Additionally, we estimate the luminosities emitted by the optically thick regions from their measured dust temperatures and sizes, assuming a spherical geometry with a constant radius at all wavelengths. This luminosity is given by

\begin{equation}
		L_{op} = 4 \pi \sigma r_{decon}^{2} T_{d}^{4},
\end{equation}

\noindent
where $r_{decon}$ is the deconvolved radius and T$_{d}$ is the adopted dust temperature. The resulting luminosities range from 0.08 - 3.61 L$_{\odot}$ (Table \ref{tbl:pbrsproperties}). HOPS~404 has the largest $L_{op}$ due to the highest dust temperature where $L_{op}$ exceeds the values of $L_{bol}$and $L_{tot}$. This may result from an underestimate in the amount of foreground extinction and the degree of beaming of the luminosity by the outflow cavity in the determination of $L_{tot}$.  Alternatively, the simple spherical, constant temperature geometry assumed by Lop may overestimate the flux. We note that in the case of HOPS~400, the SED and model fit luminosities are for the combined system, while $L_{op}$ is for HOPS~400-B alone. 

We also calculate two timescales in Table \ref{tbl:timescales}. The first is the free fall timescale, $t_{ff}$ = (3$\pi$/32G$\rho$)$^{0.5}$, which is determined assuming $\rho$ = 3M/4$\pi r_{decon}^{3}$ with the values in Table \ref{tbl:pbrsproperties}. The resulting high densities, i.e., $>$ 10$^{-13}$ g cm$^{-3}$ puts these sources at the critical point where the emission should be optically thick to their own radiation \citep{masunaga1998}. The resulting free fall timescales are all less than 150 years (Table \ref{tbl:timescales}). 

A second timescale is a cooling time and gives the rate of contraction for a given mass and radius assuming that all the heating due to contraction is radiated away. This is given by the Kelvin-Helmholtz time for luminosity, 

\begin{equation}
t_{KH} = \alpha \frac{GM^{2}}{LR},
\end{equation}

\noindent
where we assume $\alpha = 3/5$ for a constant density sphere.
Adopting L$_{op}$ results in times ranging from 250-3600 years. Adopting L$_{bol}$ from \citet{furlan2016}, the $t_{KH}$ range from 440-2200 years and adopting L$_{tot}$ yields 250-2000 years (Table \ref{tbl:timescales}). For all the irregular PBRs, except HOPS~404 the $t_{KH}$ are consistently longer than the free fall times. These values, however, depend the mass, which in turn depend strongly on the adopted opacity law (Table \ref{tbl:kappa}). We discuss the ranges of plausible values and their implications in \S \ref{sec:discussion}.

\begin{deluxetable*}{ccccc}
\tablewidth{0pt}
\tablecaption{8~mm Dust Opacity Values from Literature
 \label{tbl:kappa}}
\tablehead{
\colhead{Reference} & \colhead{Composition$^{c}$} & \colhead{Density (cm$^{-3}$)} & \colhead{Timescale (yr)} & \colhead{$\kappa_{8mm}$ (cm$^{2}$ g$^{-1}$)}} 
\startdata
Ossenkopf \& Henning (1994)$^{a}$ & MRN no ice & $10^{6}$ & $10^{5}$ & 0.2  \\
Ossenkopf \& Henning (1994)$^{a}$ & MRN no ice & $10^{8}$ & $10^{5}$ & 0.76  \\
Ossenkopf \& Henning (1994)$^{a}$ & MRN with thin ice mantles & $10^{6}$ & $10^{5}$ & 0.05 \\
Ossenkopf \& Henning (1994)$^{a}$ & MRN with thin ice mantles & $10^{8}$  & $10^{5}$ & 0.06 \\
Tobin et al. (accepted)$^{b}$  & MRN with thin ice mantles & $10^{6}$ & $10^{5}$ & 0.144  \\
Woitke et al. (2016) & see ref for details & $10^{4}$ - $10^{14}$ & $10^{5}$ - $10^{6}$ & 0.2 \\
\enddata
\tablenotetext{a}{Extrapolated to 8 mm using 1~mm and 1.3~mm points in Table 1 of Ossenkopf \& Henning (1994).
\tablenotetext{b}{Extrapolated to 8~mm using the 1.3~mm opacity of 0.899 cm$^{2}$ g$^{-1}$ with $\beta$=1 \citep[see][Tobin et al. accepted]{segura2016,tychoniec2018}.}}
\tablenotetext{c}{MRN: \citet{mathis1977}}
\end{deluxetable*}

\begin{deluxetable*}{cccccc}
\tablewidth{0pt}
\tablecaption{Timescales of Collapse and Lifetime
 \label{tbl:timescales}}
\tablehead{
\colhead{HOPS ID} & \colhead{$t_{ff}^{a,b}$ (yr)} & \colhead{$t_{KH}^{a,c}$ ($L_{bol}$) (yr)} & \colhead{$t_{KH}^{a,c}$ ($L_{tot}$) (yr)} & \colhead{$t_{KH}^{a,c}$ ($L_{op}$) (yr)} & \colhead{$t_{stat}$ (yr)}}  
\startdata
400-B & 86  & 440 & 250 & 2000 & 6000  \\ 
401 & 82 & 100 & 80 & 480 & $-$   \\ 
402 & 146 & 2200 & 2000 & 3600 & 6000  \\ 
403 & 125 & 600 & 465 & 3000 & 6000  \\ 
404 &  50 & 930 & 590 & 250 & 6000 \\  
\enddata
\tablenotetext{a}{Adopting $\kappa =0.144$~cm$^{2}$~g$^{-1}$ (see \S \ref{sec:mass})}
\tablenotetext{b}{Value scales with opacity by equation $t_{ff} = (\kappa_{8mm}/0.144~{\rm cm}^{2}~{\rm g}^{-1})^{1/2}$ $\times$ $t_{ff}$(tabulated) (see \S \ref{sec:timescales})}
\tablenotetext{c}{Value scales with opacity by equation $t_{KH} = (0.144~{\rm cm}^{2}~{\rm g}^{-1}/\kappa_{8mm})^2$ $\times$ $t_{KH}$(tabulated) (see \S \ref{sec:timescales})}
\end{deluxetable*}

\subsection{Outflow Properties}
\label{sec:outflowproperties}

\begin{figure*}
\centering
\plotone{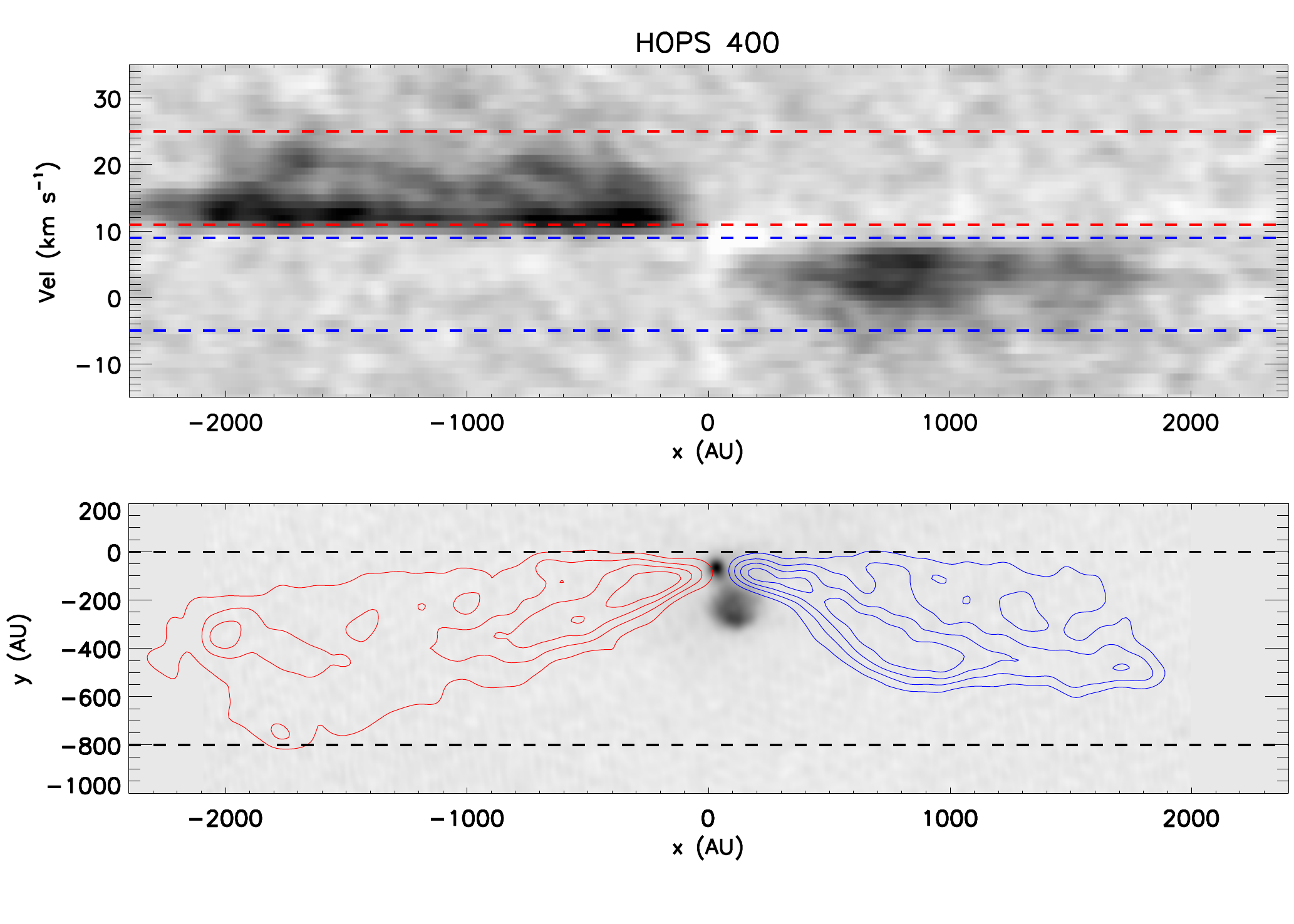}
\caption{HOPS 400 - ALMA $^{12}$CO (J=3-2)  position-velocity diagram (top) and 0.87~mm continuum image (gray-scale) with red and blue shifted integrated intensity contours at [20, 40, 60, 80] mJy beam$^{-1}$ km s$^{-1}$ (bottom). The compact outflow is centered on the continuum of 400-A and is confined to regions near the protostar. The contours are the integrated emission in the velocity ranges -5 to 9 km s$^{-1}$ for the blue lobe and 11 to 25 km s$^{-1}$ for the red lobe.  The red and blue dashed lines in the PV diagram (top) show the velocity integration range on the y-axis. The black dashed lines in the outflow contour plot (bottom) show the y-axis spatial range integrated over the PV diagram. The x-axis is at a position angle of east of north and aligned with the outflow.}
\label{fig:400outflow}
\end{figure*}

\begin{figure*}
\centering
\plotone{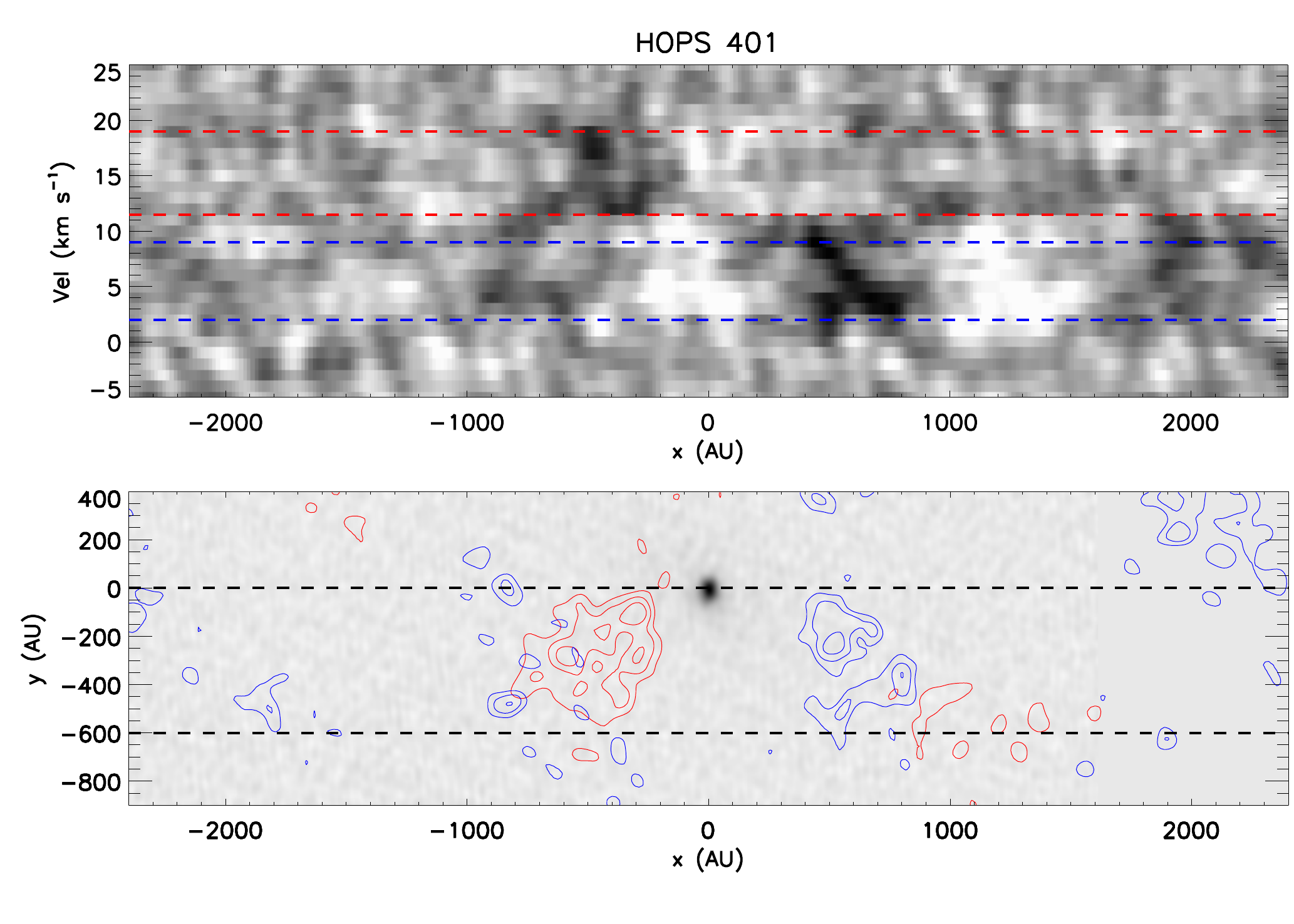}
\caption{HOPS 401 -  ALMA $^{12}$CO (J=3-2)  position-velocity diagram (top) and 0.87~mm continuum image (gray-scale) with red and blue shifted integrated intensity contours at [40, 60, 80] mJy beam$^{-1}$ km s$^{-1}$ (bottom). The contours are the integrated emission in the velocity ranges are 1 to 9 km s$^{-1}$ for the blue lobe and 11 to 19 km s$^{-1}$ for the red lobe. The red and blue dashed lines in the PV diagram (top) show the velocity integration range on the y-axis. The black dashed lines in the outflow contour plot (bottom) show the y-axis spatial range integrated over the PV diagram. A rotation angle of 315$^{\circ}$ was implemented with respect to the PA rotation of the y-axis.}
\label{fig:401outflow}
\end{figure*}

\begin{figure*}
\centering
\plotone{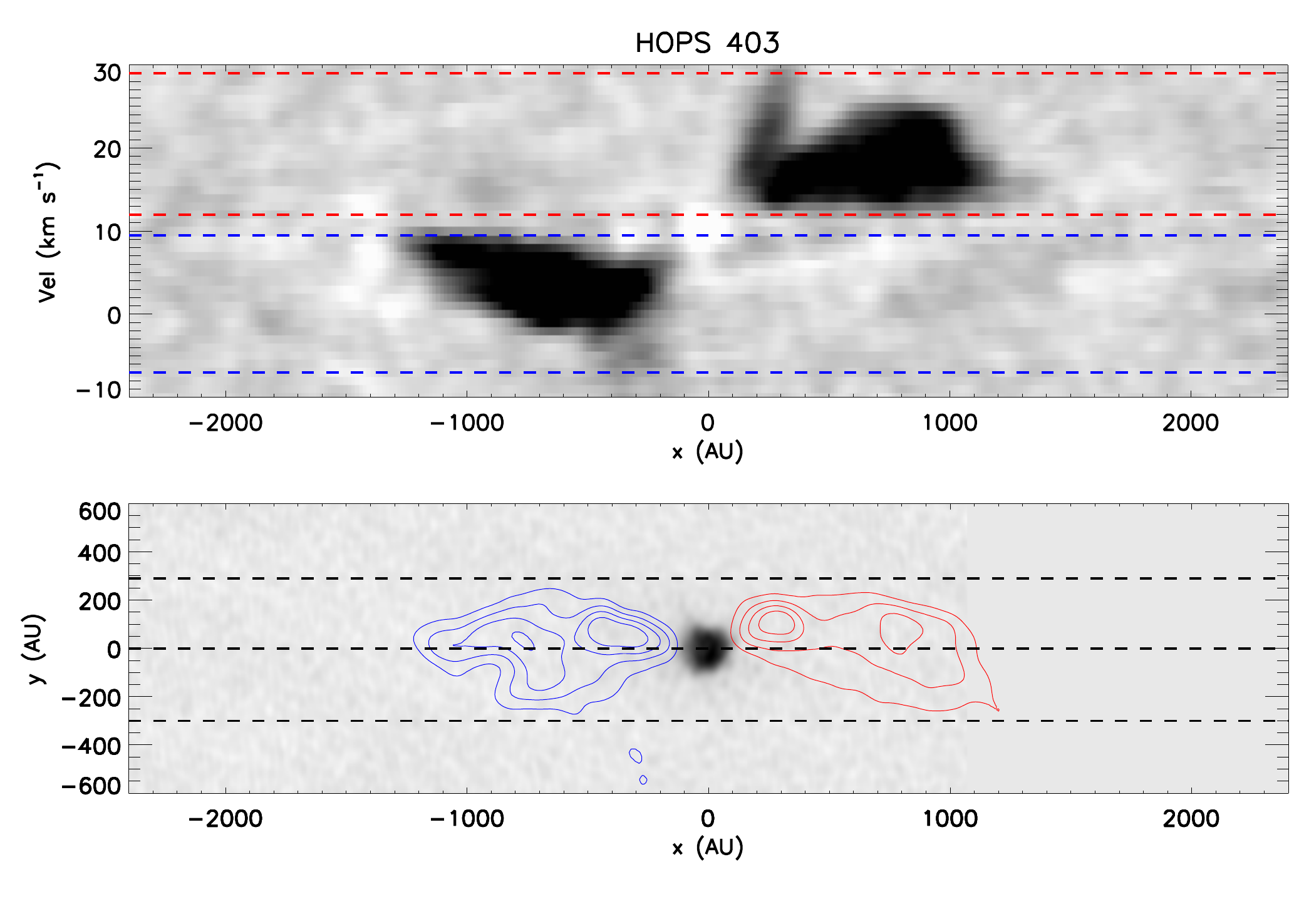}
\caption{HOPS 403 -  ALMA $^{12}$CO (J=3-2) position-velocity diagram (top) and 0.87~mm continuum image (gray-scale) with red and blue shifted integrated intensity contours at [20, 40, 60, 80] mJy beam$^{-1}$ km s$^{-1}$ (bottom). The compact outflow is centered on the continuum of HOPS~403 and is confined to regions near the protostar, but it is not detected within the continuum due to the high optical depth at 0.87~mm. The contours in the integrated intensity plot are -7 to 9.5 km s$^{-1}$ for the blue and 12 to 29 km s$^{-1}$ for the red. The red and blue dashed lines in the PV diagram (top) show the velocity integration range on the y-axis. The upper and lower black dashed lines in the outflow contour plot (bottom) show the y-axis spatial range integrated over the PV diagram. The middle dashed line is the y-axis center of the continuum image. A rotation angle of -27$^{\circ}$ was implemented with respect to the PA rotation of the y-axis.}
\label{fig:403outflow}
\end{figure*}

\begin{figure*}
\centering
\plotone{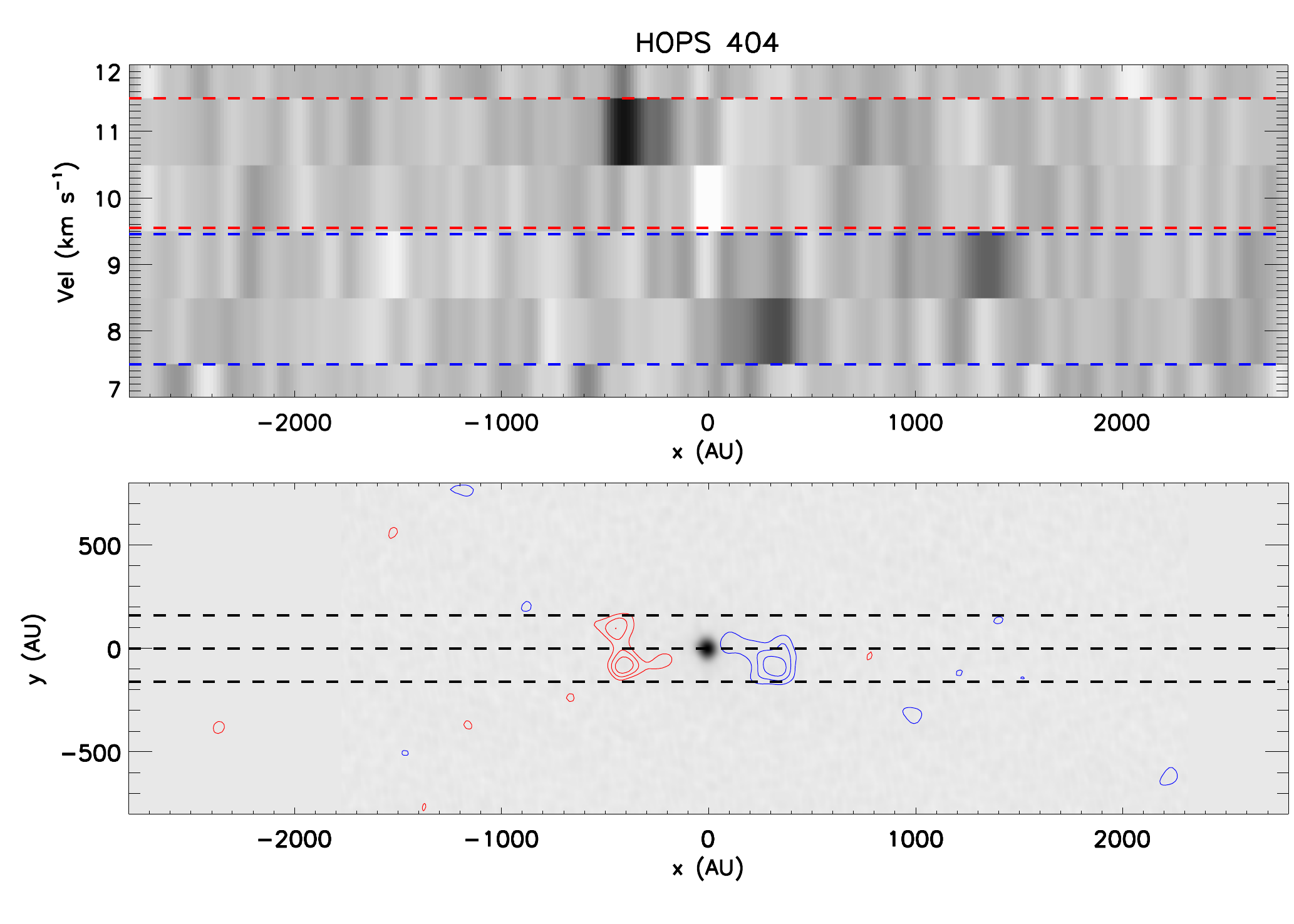}
\caption{HOPS 404 -  ALMA $^{12}$CO (J=3-2)  position-velocity diagram (top) and 0.87~mm continuum image (gray-scale) with red and blue shifted integrated intensity contours at [40, 60, 80] mJy beam$^{-1}$ km s$^{-1}$ (bottom). The compact outflow is centered on the continuum of HOPS~404 and is confined to regions near the protostar. The contours are the integrated emission in the velocity ranges are 7 to 9 km s$^{-1}$ for the blue lobe and 10 to 12 km s$^{-1}$ for the red lobe. The red and blue dashed lines in the PV diagram (top) show the velocity integration range on the y-axis. The upper and lower black dashed lines in the outflow contour plot (bottom) show the y-axis spatial range integrated over the PV diagram. The middle dashed line is the y-axis center of the continuum image. A rotation angle of 35$^{\circ}$ was implemented with respect to the PA rotation of the y-axis.}
\label{fig:404outflow}
\end{figure*}

Here we highlight the outflow properties of each source that has a detection in the $^{12}$CO (J=3-2) ALMA channel maps: HOPS~400-A, HOPS~401, HOPS~403, and HOPS~404 (Figure \ref{fig:400outflow} - \ref{fig:404outflow}). This is the first detection of an outflow from HOPS~404 and potentially HOPS~401. HOPS~400-A and HOPS~403 were previously found by CARMA observations by \citet{tobin2015,tobin2016} with HOPS~401 and HOPS~404 being non-detections at their sensitivities. 

The outflow masses are calculated using the LTE approximation following the method in \citet{yildiz2015} and assume a gas temperature of 70~K. We calculate the outflow force F$_{CO}$ using the method in \citet{vandermarel2013} assuming an inclination of 50$^{\circ}$. The systemic velocities of the PBRs are $\sim$10 km s$^{-1}$, and we use the highest velocity emission channels of each lobe in the force equation. The calculated properties of the outflows are listed in Table \ref{tbl:outflow}. We do not have inclinations for each individual outflow. Therefore, we report the projected lengths (a lower limit to the actual length) and the radial velocities (also a lower limit).  The resulting dynamical times assume a 50$^{\circ}$ inclination, with times being lower with higher inclinations and higher with lower inclinations.

The HOPS~400-A and HOPS~403 outflows have the longest lengths and the largest masses and forces in each lobe. The highest velocity emission is also in HOPS~400-A and HOPS~403. In case of HOPS~403, the continuum is opaque (Fig \ref{fig:403outflow}), and we cannot trace the outflow to its driving source. The bright, point-like source in HOPS~403 at 8~mm is the likely source of the outflow. Since HOPS~403 is optically thick, we cannot tell whether this point source is inside the opaque, irregular region or a companion located behind it. The presence of a companion behind HOPS~403 would explain why L$_{tot}$ and L$_{bol}$ of HOPS~403 are more than six times higher than the luminosity for the opaque region (Table~\ref{tbl:pbrsproperties}).

The masses of the outflow lobes in HOPS 400-A and 403 are within an order of magnitude of \citet{tobin2016}. In contrast, \citet{tobin2016} found lengths of the outflow lobes and dynamical ages 3$\times$ larger for HOPS 403 and 400-A; the ALMA data is unable to recover the large-scale outflow emission that CARMA did. The resulting values for the outflows are consistent with young outflows from either FHSCs or Class 0 protostars. We note that other outflow tracers (e.g., H$_{2}$CO) may pick up additional material that is not detected by the observations presented here and shed more light on the outflows.

In HOPS 404 (Fig \ref{fig:404outflow}), the masses of each lobe are the smallest, the lengths the shortest, the forces the lowest, and the maximum velocities are only 2 km s$^{-1}$. We discuss the implications for this in \S \ref{sec:discussion}. A more detailed description of all of the outflows is given in Appendix A.

\begin{deluxetable*}{cccccc}
\tablewidth{0pt}
\tablecaption{Outflow Properties of Sources
 \label{tbl:outflow}}
\tablehead{
\colhead{HOPS ID}  & \colhead{M$_{red}$/M$_{blue}$} & \colhead{t$_{dyn,red}$/t$_{dyn,blue}^{a}$} & \colhead{F$_{red}$/F$_{blue}$} & \colhead{R$_{red}$/R$_{blue}$} & \colhead{V$_{red}$/V$_{blue}$} \\
\colhead{$ $} & \colhead{(10$^{-3}$ M$_{\odot}$)} & \colhead{(yr)} & \colhead{(M$_{\odot}$ km s$^{-1}$ yr$^{-1}$)} & \colhead{(AU)} & \colhead{(km s $^{-1}$)}} 
\startdata
400-A &  3.2/3.0 & 825/525 & 1.95$\times$10$^{-4}$/2.38$\times$10$^{-4}$ & 2454/2018 & $+$24/-8  \\
401 & 1.6/0.26 & 360/310 & 6.44$\times$10$^{-5}$/1.02$\times$10$^{-5}$ & 680/720 & $+$19/$+$1 \\
403 & 4.1/3.9 & 316/316 & 4.40$\times$10$^{-4}$/4.46$\times$10$^{-4}$ & 1200/1200 & $+$29/-7 \\
404 & 0.02/0.03 & 1710/1105 & 1.35$\times$10$^{-7}$/2.48$\times$10$^{-7}$ & 680/483 & $+$12/$+$8 \\
\enddata
\tablenotetext{a}{t$_{dyn,red}$=R$_{red}$/(V$_{red,max}$-V$_{LSR}$), t$_{dyn,blue}$=R$_{blue}$/(V$_{blue,max}$-V$_{LSR}$), where R$_{red}$ and R$_{blue}$ are the maximum detected extend from the central source.}
\end{deluxetable*}

\section{Discussion}
\label{sec:discussion}

We have resolved four irregular protostars that are optically thick at 0.87 mm over $\sim$100~AU diameter regions, each showing peaks which may be optically thick at 8~mm within $\sim$50~AU diameter regions. We interpret these as the centers of collapsing molecular fragments that are optically thick at their peak wavelength of $\sim30$~$\mu$m, which corresponds to a temperature of $\sim$100~K.  Models of collapsing fragments show that due to the very high densities of gas in these opaque regions, $\sim 10^{-13}$~gm~cm$^{-3}$, the heating generated by compression exceeds the rate of cooling by radiation \citep{masunaga1998}. Within this thermal opacity horizon, the collapsing gas is no longer isothermal and the ensuing increase in temperature and pressure will slow and eventually halt collapse. The high optical depth, and the resulting adiabatic equation of state, should also resist fragmentation despite the presence of multiple Jeans masses within each region \citep{rees1976,low1976,silk1977}. In this way, the formation of an optically thick zone is the prelude to the formation of a HSC, and these objects may trace the earliest observable state of protostar formation. While the ALMA 0.87~mm data measure the sizes and temperatures of the opaque horizons, the VLA data, due to the lower optical depth at 8 mm, resolve the structures that are opaque at shorter wavelengths.  These data show irregular, non-circularly symmetric structures with bright, compact peaks. Although we refer to these objects as protostars, we do not know if they all contain HSCs; the presence of such a central core is often implied in the standard usage of the term protostar.  

Analogous sources have been identified in other nearby star forming regions. \citet{hernandezgomez2019} find that source B in the binary Class 0 protostar IRAS 16293-2422 has a power-law spectrum ($S_{\nu} \propto \nu^{2.28}$) from 0.5~mm to 10~cm. They argue that the observations are tracing dust emission out to 10~cm and that the dust is becoming optically thick at the shorter wavelengths.  In the Perseus molecular cloud, \citet{cox2015} find that source A1 of the NGC 1333 IRAS 4A system has an optically thick inner core and \citet{sahu2019} detect complex organic molecules in absorption against the continuum. The sample presented in this paper is the result of the first systematic search for such extended, irregular optically thick regions utilizing the large population found in Orion, the well-characterized SEDs of the HOPS program, and the high resolution ALMA and VLA maps from the VANDAM Orion survey.

Simulations of fragment collapse predict the formation of optically thick regions that are i) several AU in size, ii) thermally supported, iii) and with masses $< 0.1$~M$_{\odot}$ \citep[e.g.,][]{bhandare2018}. These objects are the FHSCs first described by \citet{larson1969}. Only simulations invoking large amounts of rotational energy, resulting in the formation of highly flattened pre-stellar disks instead of spherical HSCs, predict masses and radii comparable to those of the envelopes \citep{bate2011, commercon2012}. As we will discuss in the remainder in this section, there are several areas of tension between the observed results and existing models. 

\subsection{The Statistical Lifetime}
\label{sec:stat}

\begin{figure}[t]
    \begin{minipage}[t]{0.99\linewidth}
        \includegraphics[width=\linewidth]{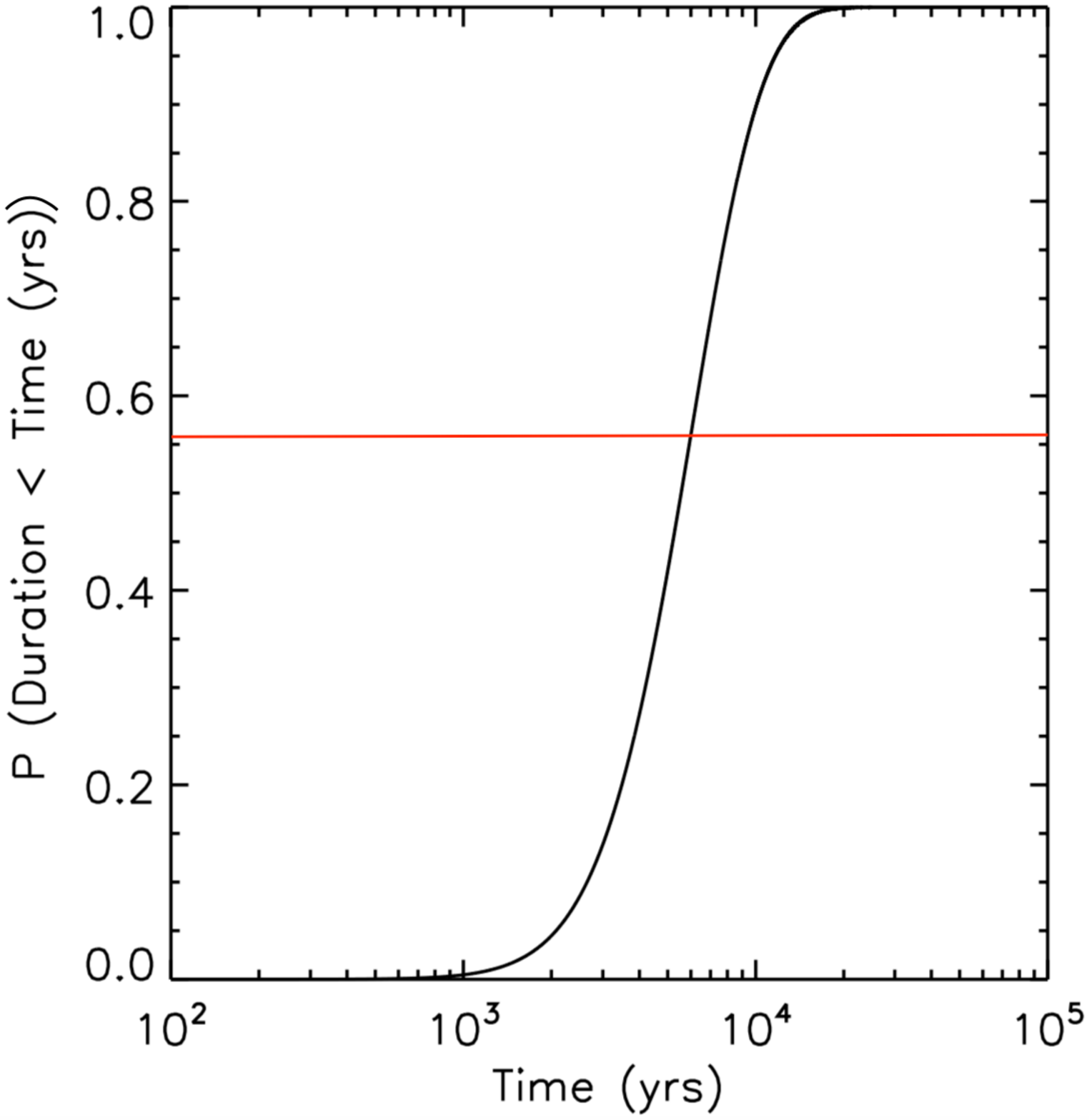}
\caption{The probability of detecting four optically thick protostars vs.~the lifetime of the optically thick phase. Each value of the probability corresponds to a different lifetime. The lifetime of the optically thick phase is the number of optically thick sources observed divided by the total number of protostars and multiplied by the total protostellar lifetime of 0.5 Myr \citep{dunham2014}. The red horizontal line indicates the probability of the protostars having a mean lifetime of 6000 years or less.}
\label{fig:stat}
\end{minipage}
\end{figure}

The detection of only four irregular, optically thick protostars out of a sample of 328 protostars places constraints on the lifetime of such objects, i.e., t$_{stat}$. If we assume i.) that the formation of extended, opaque zones is an early stage all protostars go through and ii.) that the star formation rate is constant over a protostellar lifetime, the duration of this stage can be estimated from the ratio of the number of extended sources over the total number of protostars. For a protostellar lifetime of 0.5 Myr \citep{dunham2014}, the duration is t$_{stat} = 4/328~ \times 0.5~{\rm Myr} \approx 6000$~yrs. It is possible that the assumptions in $t_{stat}$ are incorrect, e.g., \citet{kristensen2018}. Using the binomial theorem, we determine the cumulative probability distribution for finding four optically thick sources as a function of the lifetime of the optically thick stage (Figure \ref{fig:stat}). This figure shows the distribution of probabilities as a function of lifetime and demonstrates that the lifetimes under 1000 years have very low chances ($<$1\%) of producing four distinct objects within the Orion sample. The optically thick stage length of 6000~yrs is comparable to the $t_{KH}$ lifetime, particularly when adopting the optically thick region luminosities ($L_{op}$) in \S ~\ref{sec:timescales}.

\subsection{Inside the Opacity Horizon}
\label{sec:energy}
If these objects are tracing the transition of a collapsing opaque region to protostar, the structure, kinematics, and thermodynamics of the gas within the opaque zones are essential for understanding the conditions that lead to the formation of the HSCs and guide their early ($< 10,000$~yrs) evolution. Due to the high optical depths and limited sensitivity in the 0.87~mm band, we do not detect molecular lines that can be used to directly observe the motions of the gas. Instead, we must infer the possible state of the gas from the masses, radii, and luminosities. 

\begin{figure*}
\centering
\plotone{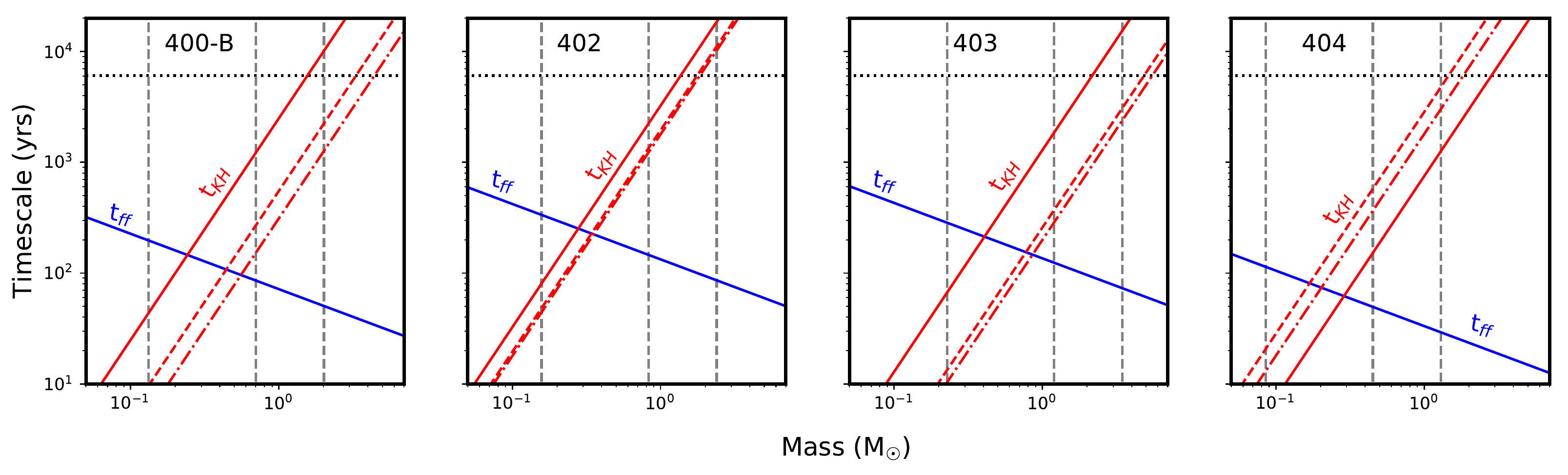}
\caption{The timescale for collapse vs.~mass for HOPS 400-B, HOPS 402, HOPS 403, and HOPS~404. The blue line is the $t_{ff}$ while the red lines are $t_{KH}$ for the $L_{op}$ (solid), $L_{bol}$ (dashed) and $L_{tot}$ (dot-dashed). The horizontal dotted line is the $t_{stat}$ = 6000 yrs, while the vertical lines give the minimum, fiducial, and maximum masses based on the opacity laws in Table \ref{tbl:kappa}.}
\label{fig:timescales}
\end{figure*}

To do this, we compare $t_{ff}$, $t_{KH}$, and $t_{stat}$. Both $t_{ff}$ and $t_{KH}$ are a function of the mass and radius, while $t_{KH}$ is also dependent on the luminosity (see footnotes of Table~\ref{tbl:timescales}). The value of $t_{stat}$ is independent of the derived physical parameters. We show the dependence of the three timescales on mass, which varies with the adopted dust opacity, in Figure~\ref{fig:timescales}. Here we find that there are no masses where all three timescales are equal. Low masses are needed for $t_{ff} \approx t_{KH}$, while high masses are needed for $t_{KH} \approx t_{stat}$. 

\begin{figure*}
\centering
\plotone{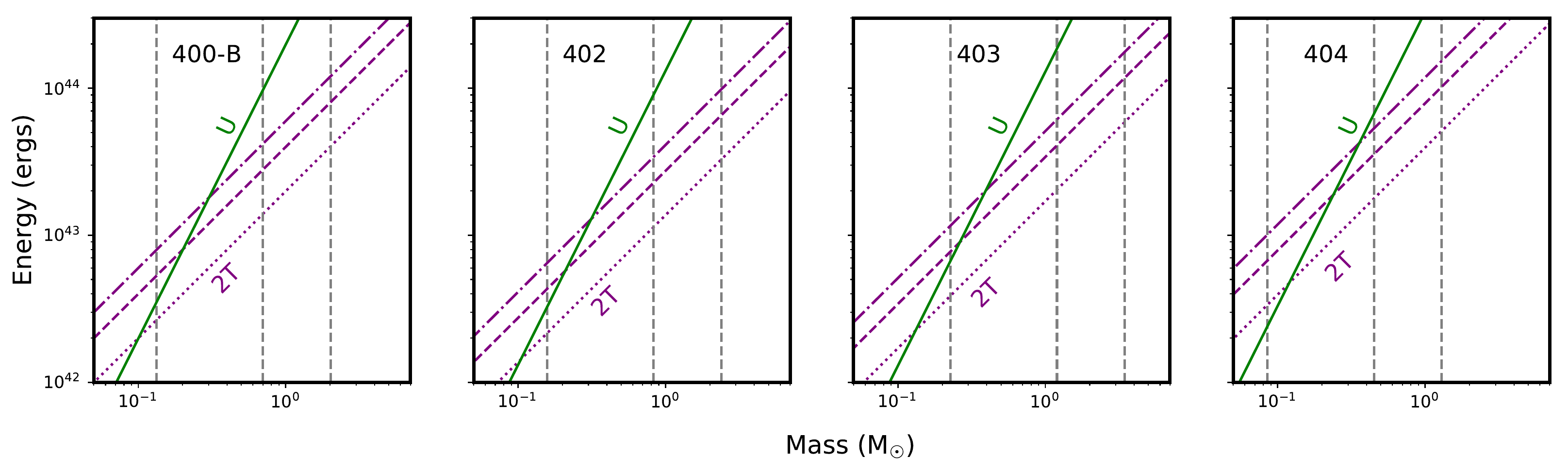}
\caption{The energy vs.~mass for HOPS 400-B, HOPS 402, HOPS 403, and HOPS~404. The green line is the potential energy $U$ while the purple lines are multiples of the kinetic energy $T$, specifically $2T$ (dotted), $4T$ (dashed), and $6T$ (dot-dashed). These represent the twice thermal energy alone, twice the thermal and turbulent energy in equipartition, and twice the thermal, turbulent and magnetic energy in equipartition. In virial equilibrium, these lines would intersection the line for $U$. As in Figure \ref{fig:timescales}, the vertical lines give the minimum, fiducial, and maximum masses for the range of dust opacities.}
\label{fig:energies}
\end{figure*}

There are no plausible masses where $t_{ff} \approx t_{stat}$. This suggests that the collapse is not on a free fall timescale. Indeed for higher masses we find that $t_{KH} \approx t_{stat}$, while both are an order of magnitude higher than $t_{ff}$. This would suggest that the opaque zones may not be in free fall but are slowly contracting as they radiate away their thermal energy. The difficulty with this interpretation is that the thermal pressure is insufficient to support the opaque zones against collapse.  This is demonstrated in Figure~\ref{fig:energies} and the last column of Table \ref{tbl:pbrsproperties}, where we compare the gravitational potential energy, $U$, to the kinetic energy, $T$, as a function of the adopted mass.  Here we assume the opaque zones can be approximated as constant density spheres with the radii given in Table~\ref{tbl:pbrsproperties} and no external pressure. We calculate the thermal energy using the temperatures  in Table~\ref{tbl:pbrsproperties}. The zones are in virial equilibrium when $2T = U$. We also plot curves equal to $4T$ and $6T$, these would represent cases where there is equipartition between the thermal and turbulent energies or equipartition between the thermal, turbulent and magnetic energies, respectively. In none of the three cases do we find a mass where the opaque regions can be in virial equilibrium and t$_{stat}$ can be satisfied. Given the lack of a consistent model, we consider four different scenarios below.

\subsubsection{The Gas is in Free Fall}

The simplest scenario is that the opaque regions are in free fall with lifetimes of only a few hundred years. In the case that the masses are at the low end of the plausible mass range, t$_{ff}$ and t$_{KH}$ can be similar. These smaller masses require either large grains to increase the dust opacity or temperatures that exceed the measured values by a factor of a few.  If the former is correct, the dust opacity must be larger than the values of 0.144 cm$^{2}$ g$^{-1}$ at 8~mm and $\beta=1$ adopted for the fiducial masses given in Table~\ref{tbl:pbrsproperties}. This opacity already attempts to account for a significant degree of grain growth over a $\beta$=1.8 model. Thus, substantial additional grain growth would need to happen in the starless fragment stage prior to the onset of the HSC formation. In order for grains to grow to mm size very high densities over 100,000 year timescales need to be maintained \citep[Table~\ref{tbl:kappa};][]{ossenkopf1994,ormel2011}, and it is not clear that the presence of such large grains is realistic. The presence of such large grains would also imply that the regions where the 0.87~mm emission is optically thick are smaller than shown in Figs~\ref{fig:hopsimages} \& \ref{fig:temps}, which may have detectable consequences, particularly for future molecular line observation. Future extensions of the continuum measurements to longer wavelengths will also put additional constraints on grain size distribution in these sources.

Alternatively, the gas may be in free fall but  $t_{KH} > t_{ff}$; in this case the opaque zones will rapidly heat up as they are compressed until the increasing central pressure is able to halt or slow collapse. In either case, the implied lifetimes are not consistent with $t_{stat}$, and either this is a very unlikely coincidence, the statistical probability of observing four such regions in free fall is $<$1\%, or $t_{stat}$ is not representative of the lifetimes of these sources. 



\subsubsection{Turbulence Resists Collapse}
\label{sec:turbulence}

If the statistical lifetime is correct, then these opaque regions cannot be in a state of free fall. In this case, the contraction times implied by the measured luminosities require masses toward the higher end of the mass range. Furthermore, additional support is required to keep the large masses within the opacity horizon from collapsing or to slow down the rate of collapse. 

We first consider whether turbulence can supply that support. This can be quantified by considering the virial theorem for the central opaque zones \citep{mckee1999,hartmann2009}

\begin{equation} 
 3\frac{M kT}{\mu m_H} + 3 M \sigma_{turb}^2 + W =  \frac{3}{5}\frac{GM^2}{R} + \int_S P~\vec r \cdot \vec dS, 
 \label{eqn:virial}
\end{equation}

\noindent
where $\sigma_{turb}$ is the turbulent linewidth, $R$ is the radius of the opaque zone, $M$ is the enclosed mass, $P_{S}$ is the surface pressure at the opacity horizon, and $W$ is the magnetic term that we will discuss in the following scenario. Ignoring the surface pressure and magnetic terms, we estimate that turbulent velocities of $\sim$1.3~km~s$^{-1}$ are required, assuming virial equilibrium and adopting the radius and mass in Table~\ref{tbl:pbrsproperties} for HOPS~402. Compared to the sound speed of 0.34~km~s$^{-1}$, the velocities are supersonic. These motions would dissipate on a crossing time \citep{maclow1998}, $t_{cross} = {R}/{\sigma_{turb}}$, where $t_{cross}$ = 300~yrs for HOPS~402, and the turbulent support would dampen in a few hundred years.  

One possible way of sustaining turbulence may be the input of energy by infall from the surrounding envelope of gas. \citet{klessen2010} propose that accretion may be a universal mechanism for driving turbulence. The linewidth that can be sustained is given by 

\begin{equation}
\sigma_{turb} = \left(2 \epsilon R v^2_{infall} \frac{\dot M}{M}\right)^{1/3},
\end{equation}

\noindent
where $v_{infall}$ is the infall rate and $\epsilon$ is an efficiency factor that we set to 0.1 \citep{heitsch2013}. For an infall rate of 10$^{-4}$~M$_{\odot}$~yr$^{-1}$ \citep{stutz2013,furlan2016}, the resulting linewidth is $\sigma_{turb}$ = 0.6~km~s$^{-1}$. This shows that the estimated infall rate, although high for low mass protostars, is insufficient to drive the required level of supersonic turbulence. 
For the protostars with outflows or companions with outflows, it is possible that the outflow is driving the turbulence \citep{offner2018}, however, for protostars without outflows it is unlikely that the turbulence can be sustained for more than a few hundred years.    

\subsubsection{Magnetic Fields Help Resist Collapse}
\label{sec:magnetic}
An alternative form of support can be  magnetic fields. In equation \ref{eqn:virial}, this term is given as 

\begin{equation}
W = \int_V \frac{B^2}{8\pi}dV +\int_S \vec r \cdot \left(\vec B \vec B - \frac{1}{2}B^2 {\bf I}\right)\cdot \vec dS,
\end{equation}

\noindent
where $I$ is the unit tensor.  As discussed in \citet{hartmann2009}, the collapse of a gas fragment requires that the magnetic term in the virial theorem be less than the potential energy term ($U$) since both the magnetic and potential energy terms scale approximately as $1/R$ as the fragment shrinks, assuming the magnetic field is frozen to the gas.

It has been found, though, that the magnetic energy can exceed thermal and turbulent energies. Magnetic field energies that dominate the turbulent and kinetic energies, but not the potential energy, have been found for the high mass protostars G9.62+0.19 and G31.41+0.31 \citep{girart2009,dallolio2019}. A magnetic field that is dominated by gravity, but not turbulent motions, can produce the hourglass shaped magnetic fields observed toward low mass protostars and starless cores \citep{myers2018}.  Furthermore, on large scales, \citet{stutz2016} found that in the Integral Shaped Filament (ISF) in Orion A, the potential energy dominates on large scales, while magnetic energy density dominates on small scales; moreover, in subsequent work, \citet{gonzalez2019} show that the turbulent energy is insufficient to support the ISF. 

If the magnetic field dominates over the total kinetic energy, but not the gravitational energy, the increase in temperature at high densities will boost the thermal energy and, by raising the sound speed,  also elevate the sub/trans-sonic turbulent motions.  This boost in thermal and turbulent motions  increases the support against collapse. In this way, it is possible that the support of the opaque zone can be primarily - but not solely - due to the magnetic field.  One caveat is that strong magnetic fields often lead to flattened structures in models \citep[e.g.,][]{commercon2012}, this is inconsistent with the observed morphologies of the opaque zones, unless seen face-on (see next scenario).

\subsubsection{Rotational Support and Disks}

Rotational motion can also provide support against collapse in cases where the rotational energy is high, producing extended disks \citep{bate1998,saigo2008}.  \citet{bate2011} simulated the collapse of a 1~M$_{\odot}$, uniform density, uniformly rotating sphere with rotational to gravitational energy ratios ranging from 0 to 0.04. While the low value resulted in a few AU diameter, hydrostatically supported FSHC, the larger values led to rotationally supported pre-stellar disks that extend to radii of 150 AU, have masses approaching 0.2~M$_{\odot}$, and lifetimes extending to 3000~yrs \citep[Figure 15 from][]{bate2011}. The temperatures and densities of the simulated pre-stellar disks are also similar to those of the irregular protostars. The irregular morphologies may arise from gravitational instabilities in the disks.  

The primary difficulty with rotational support is that models invoking rotation predict highly flattened morphologies. The aspect ratios of the irregular structures are close to 1, particularly when considering the outer contours of the 0.87 mm emission. This suggests that they would need to be observed at a nearly face-on inclination if they were pre-stellar disks. It is possible (although quite unlikely) that all four irregular PBRs are observed at this inclination, but there are other difficulties with pure rotational support. In particular, the orientation of the outflows in HOPS 404 and 403, if not a binary, would rule out a face-on orientations if the outflows are oriented perpendicular to the disks. 

Although disks remain a plausible explanation, it is very unlikely that the irregular protostars are stable Keplerian disks surrounding protostars, i.e, young Class 0 protostars with massive disks. Again, the aspect ratios $\sim 1$ for the observed disks would imply that the disks are observed at nearly face-on inclinations. At such inclinations, the SEDs would be less red than typical Class~0 protostars due to the lower extinction along the angular momentum axis \citep{furlan2016}. In contrast, the PBRs are distinguished by their low $T_{bol}$ and weak mid-IR emission compared to other Orion Class 0 protostars \citep{stutz2013}. Furthermore, the red and blue outflow lobes of a face-on protostar would overlap, which is not the case for the observed outflows.  Finally, the irregular sources do not look like the disks resolved around many protostars in the VANDAM sample (Tobin et al. accepted).

As we will discuss in the next section, it is likely that one to two of the opaque regions, HOPS~404 and potentially HOPS-403, contains HSCs. These HSCs may have formed in  pre-stellar disks \citep[e.g. ][]{bate2011} or, perhaps, formed before the disks. In both cases, the irregular structures, high  masses, and off centered peaks observed at 8~mm suggest that the disks would have masses similar to or higher than those of the HSCs.  The opaque regions may also contain smaller, more stable disks; these could be responsible for the observed outflows.

\subsection{Outflows as Tracers of Hydrostatic Cores}
\label{sec:outflow}

Although fragmentation of the high opacity, adiabatic gas may be suppressed compared to that of the isothermal gas \citep[e.g.,][]{silk1977}, models show the formation of denser objects within the high opacity zones. In pre-stellar disks, gravitational instabilities lead to the formation of multiple HSCs \citep{bate2011}, although this fragmentation can be suppressed by magnetic fields \citet{commercon2012}. Even for hydrostatically supported FHSCs, the dissociation of H$_{2}$ in the centers of the HSCs leads to the formation of second hydrostatic cores (i.e., protostars) inside the FHSCs \citep[e.g.][]{larson1969,basu1994}.

Within the observed opacity horizons, there is evidence that HSCs have formed. The VLA maps show compact peaks that are potentially optically thick at 8~mm and could contain HSCs. Furthermore, the presence of outflows from HOPS 404, and posibly from HOPS~403, suggest the presence of HSCs (i.e., young stellar objects) whose gravity is needed to generate outflows.  

The outflow properties derived in \S \ref{sec:outflowproperties} place constraints on the mass and radius of the objects driving the outflows. The outflows are likely driven by the coupling of magnetic fields to disk rotation, and therefore the maximum velocity of the outflow is proportional to the escape velocity from  the HSC,

\begin{equation}
v_{out} = \frac{R_a}{R_0} v_{esc}
\end{equation}

\noindent
\citep[Equation 12 from][]{pudritz2007}, where $R_0$ is the minimum radius at which the magnetic field lines thread  the disk.  The escape velocity at $R_0$  is given by

\begin{equation}
	v_{esc} =  \left(\frac{2 GM }{ R_0 }\right)^\frac{1}{2},
\end{equation}

\noindent
where M is the mass of the HSC. This assumes that the mass is dominated by the HSC. The central temperature of the HSC is given by

\begin{equation}
    T_c = \alpha \left(\frac{\mu m_H}{k}\right)  \left(\frac{GM}{R}\right),
\end{equation}

\noindent
where $R$ is the radius of the HSC. Assuming that $R_0 \approx 4 R$, which is consistent with the inner disk radii inferred for young stellar objects, \citep[e.g.][]{bouvier2007}, we can write the central temperature of the HSC as

\begin{equation}
	T_{c} \approx \alpha \left(\frac{2 \mu m_{H}}{k}\right)\left(\frac{R_0}{R_{a}}v_{out}\right)^2.
\end{equation}
	
\noindent
In this equation, $R_0$ is the radius of the HSC and $R_{a}$ is the launching radius of the outflow. To evaluate the equation, we assume a convective star (i.e., a n=1.5 polytrope) where  $\alpha = 0.54$,  m$_{H}$ is the mass of a hydrogen atom, and $\mu$ is the mean molecular weight, which is 2.4  for  molecular gas and 1.33 for atomic gas. The ratio of $R_{0}$ to $R_{a}$ is taken as 1/3 \citep{pudritz2007}. 

The maximum velocity of $\sim$18 km s$^{-1}$ for each outflow lobe in HOPS~400-A and HOPS~403, yields a central temperatures of $\sim$10,000~K, which is above the dissociation temperature of molecular hydrogen. The actual maximum velocities and resulting temperatures may be a factor of $2$ higher due to the inclination of the outflow. The short dynamical ages of the outflows support the idea that these are very young protostars that have evolved past the FHSC stage.
\
\begin{figure*}
\centering
\plotone{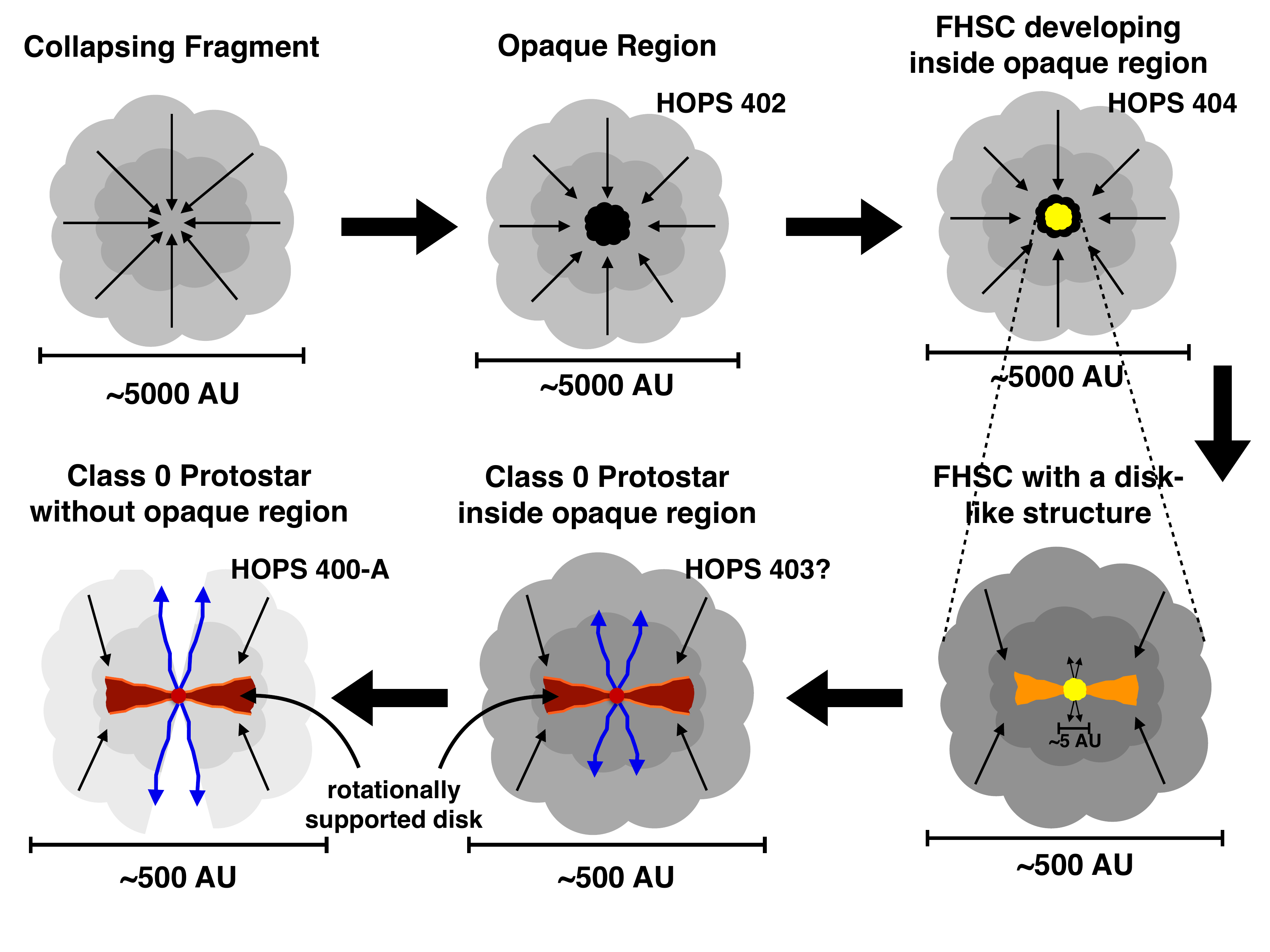}
\caption{ A cartoon schematic of the evolutionary sequence for the earliest phases of low-mass star formation, as suggested by our observations and as seen from an edge-on inclination. The upper left begins with a collapsing fragment of gas and dust. The upper middle continues with the formation of an opaque region at the center of the collapsing fragment with continuing gravitational infall. The upper right is a FHSC forming within the opaque region. The bottom right is a FHSC with a disk-like structure and an outflow beginning. The bottom middle depicts the formation of a Class 0 protostar inside the opaque region that may have a rotationally supported disk and more well-defined outflows. The final step depicted in the bottom left of a typical Class 0 protostar with outflows that have broken through the envelope, an actively accreting, rotationally supported disk, and continued infall from the slowly dissipating envelope.}
\label{fig:cartoon}
\end{figure*}
\\

On the other hand, the maximum outflow velocity of HOPS 404 is $\sim$2 km s$^{-1}$ in both outflow lobes. This implies a central temperature of $\sim$140~K, which is below the dissociation temperature of molecular hydrogen by more than an order of magnitude. Thus, the slow velocity suggests the outflow is driven by a FHSC. This FSHC would be unresolved and located within the extended dust emission that we observe. 

The maximum velocities of 2 km s$^{-1}$ are also consistent with model predictions of FHSC outflow velocities \citep{price2012}. In comparison, \citet{gerin2015} detect outflows associated with the Barnard 1b FHSC candidate with velocities up to $\sim$7~km s$^{-1}$ and estimated dynamical age of $\sim$1000 and $\sim$2000~years for B1b-S and B1b-N, respectively. They note that the outflow masses, mass-loss rate, and mechanical luminosities agree with theoretical predictions of FHSC.

The presence of HSCs in HOPS~400-B and HOPS~402 are less clear and the data presented here can not absolutely rule out outflows from either source. The 8~mm continuum image of HOPS~400-B demonstrates a compact emission peak off center from the overall emission that may mark the presence of a hydrostatic core. HOPS~402 lacks a clear compact emission peak in either wavelength and the brightness temperature profiles are the broadest and coldest in this sample (Fig. \ref{fig:temps}). 

\subsection{Evolutionary Context and Scenario for Irregular PBRs}
\label{sec:cartoon}
Figure \ref{fig:cartoon} is a cartoon schematic visualizing the formation of a low-mass star as we interpret in this paper, incorporating our interpretation of how the irregular PBRs fit into the broader context of star formation. The viewing angle is assumed to be edge-on with respect to the angular momentum axis. In the upper left, the first step is a collapsing fragment of gas and dust. An opaque region develops in the central region of the collapsing fragment, which is represented in the top middle. The top right represents a FHSC developing inside the opaque region with a disk-like structure forming. The bottom right shows the inner region, where a rotationally supported disk or disk-like structure has formed, as well as an outflow. The bottom middle is a Class 0 protostar inside an optically thick region that contains a rotationally supported disk and more well-defined outflows. Finally, depicted in the bottom left, the Class 0 continues to evolve without an optically thick region. This phase has an outflow driven by a rotationally supported disk.

We interpret HOPS~402 as an opaque region that may not have a FHSC. HOPS~400-B may be more evolved than HOPS~402 due to the point-like peak at 8~mm, but still may not have a HSC. HOPS~403 may represent an intermediate phase where a Class 0 protostar is launching an outflow but it still has a large opaue region, if it is not a binary. The Perseus protostar IRAS 4A1 \citep[scenario II of][]{sahu2019,santangelo2015} may also be an example of this phase. HOPS~404 is a FHSC candidate that has developed inside an optically thick region. The outflow detected implies a disk or disk-like structure within the opaque region. Once the central temperature of the FHSC reaches $\sim$2000~K it evolves into a Class 0 protostar with a rotationally supported disk and more well-defined outflows. HOPS~403 may represent an intermediate phase where a Class 0 protostar is launching an outflow but it still has a large opaque region. HOPS~400-A appears to be a young Class 0 protostar with a small disk and more typical flux densities at 0.87~mm and 8~mm.



\subsection{The Concentration of Irregular PBRs in Orion~B}

As noted by \citet{stutz2013}, 13 of the 19 PBRs are located in the Orion B cloud \footnote{One source previously identified as being in the Orion~B cloud, HOPS~354, is found in the Lynds~1622 cloud.  An analysis of Gaia DR2 parallaxes by \citet{kounkel2018} indicates that this cloud is in the foreground of Orion~B and is the same distance as the Orion OB1a association (345$\pm$6~pc).}. This concentration is even more extreme for the irregular PBRs, all of which are found in Orion~B. In contrast, Orion~B has 69 out of 330 protostars and 30 out of the 62 Class 0 protostars in the HOPS survey. The concentration of irregular PBRs in Orion~B may be explained by environmental differences in Orion~B, which has led to the formation of the irregular PBRs. In this case, $t_{stat}$ would be significantly higher since in our original calculation, we assumed the total number of protostars in Orion A and B.  Alternatively, Orion~B may be undergoing a rapid acceleration in the star formation rate \citep{fischer2017}. Either of these proposed solutions has important implications for the star formation process, and future observations and modeling are needed to further understand the concentration of PBRs in Orion. 

\section{Summary}
\label{sec:summary}
We present the detection of four extended, irregular structures resolved in $\sim$0.1$\arcsec$ (40~AU) resolution ALMA 0.87~mm and VLA 8~mm imaging that may represent the youngest protostars in Orion. These protostars are HOPS 400-B, 402, 403, \& 404 \citep{furlan2016}, and are part of a sample of very red, young protostars identified in $Herschel$ observations of Orion \citep[the PACS Bright Red sources or PBRs]{stutz2013}.

The four irregular protostars are distinguished by high VLA flux densities compared to the remainder of the PBRs sample, and by their bright 0.87~mm emission that is optically thick over most of the observed structures. This emission extends out to radii of $\sim$55 - 84~AU (half maximum) and implies peak dust temperatures ranging from 41 - 170~K. Based on their average dust temperatures, the masses estimated from the 8~mm data range from 0.5 to 1.2~M$_{\odot}$ and depend strongly on the adopted opacities. These appear to be the center of collapsing cloud fragments that are optically to their own radiation.  

There are three timescales associated with the opaque regions of these protostars: the free fall time ($t_{ff}$), the Kelvin-Helmholtz time ($t_{KH}$) or cooling time, and the statistical lifetime ($t_{stat}$).  The $t_{ff}$ for the four irregular protostars are $\sim$100 years, whereas, the $t_{KH}$ determined with the observed luminosities range from 250 to 3600 years. The masses used to calculate $t_{KH}$ strongly depend on the adopted dust opacities. The value of $t_{KH}$ can be equal to the $t_{ff}$ for masses at the low end of the plausible range. However, since the irregular PBRs make up $\sim$1.5\% of the HOPS protostars, their statistical lifetime is 6000 years if they represent a distinct phase in protostellar evolution, assuming constant star formation rate. This is close to the higher values of $t_{KH}$ obtained with our higher estimated masses and requires internal support against gravity to prevent collapse. Due to their consistence with $t_{stat}$, we favor the longer times and higher masses. 

Thermal pressure alone is insufficient to support the irregular PBRs from collapse, and alternative means of support are required.  These include rotational support or a combination of magnetic, turbulent, and thermal support with the magnetic field dominating. 

In the case of rotational support, the objects may be highly unstable disks, such as the pre-stellar disks predicted by simulations. The irregular structure and the lack of mid-IR emission or outflows expected for the face-on inclination implied by their morphologies make it unlikely that these are stable disks surrounding young protostars.


The compact outflows we detect in $^{12}$CO (J=3-2) line indicate the presence of HSCs toward three sources with minimum ages set by the dynamical times of $\sim$300 - 1400 years.  One of the sources is HOPS 400-A, a companion Class 0 protostar to the irregular HOPS 400-B. The other is located toward a peak of HOPS 403, this may be from a source within the opaque region or a companion behind it. Finally, the low velocity outflow from HOPS~404 suggests that it contains a FHSC.


Future studies of these irregular PBRs should include polarimetry designed to resolved the magnetic field structure and molecular line width observations to measure turbulent and/or rotational motions of the inner envelopes. These will reveal whether magnetic fields, rotation and/or turbulent motions are sufficient to support the opaque regions against collapse.

\vspace{12 pt}
\indent \textbf{Acknowledgements}
\\
We wish to thank the anonymous referee for insightful comments and suggestions that improved the quality of this manuscript. The authors acknowledge Lee Hartmann for the useful discussions and many insightful comments.
Support for Nicole Karnath was provided by the NSF through award SOSP AST-1519126 from the NRAO and by the NASA Origin of the Solar System program 13-OSS13-0094 (Megeath PI for both). This work made use of the SIMBAD database, the Vizier database, and the NASA Astrophysics Data System, funded by the National Aeronautics and Space Administration and the NSF.

AS acknowledges funding through Fondecyt regular (project code
1180350), ``Concurso Proyectos Internacionales de Investigaci\'on''
(project code PII20150171), and Chilean Centro de Excelencia en
Astrof\'isica y Tecnolog\'ias Afines (CATA) BASAL grant AFB-170002.

JJT acknowledges funding from the National Science Foundation AST-1814762.

ZYL is supported in part by NSF 1716259 and  NASA 80NSSC18K1095. 

GA, MO, and AKD-R acknowledge financial support from the State Agency for Research of the Spanish MCIU through the AYA2017-84390-C2-1-R grant (co-funded by FEDER) and through the ``Center of Excellence Severo Ochoa" award for the Instituto de Astrofisica de Andalucia (SEV-2017-0709).

This research was conducted in part at the SOFIA Science Center, which is operated by the Universities Space Research Association under contract NNA17BF53C with the National Aeronautics and Space Administration.

The National Radio Astronomy Observatory is a facility of the National Science Foundation operated under cooperative agreement by Associated Universities, Inc. This paper makes use of the following ALMA data: ADS/JAO.ALMA\# 2015.1.00041.S. ALMA is a partnership of ESO (representing its member states), NSF (USA) and NINS (Japan), together with NRC (Canada), MOST and ASIAA (Taiwan), and KASI (Republic of Korea), in cooperation with the Republic of Chile. The Joint ALMA Observatory is operated by ESO, AUI/NRAO and NAOJ.

\textit{Software}: CASA \citep{mcmullin2007}.

\section{Appendix A}
\label{sec:appendix}

\subsection{HOPS~400-A}

HOPS~400-A is a Class 0 protostar driving an outflow and is more evolved than its extended companion, HOPS~400-B. It has 0.87~mm and 8~mm flux densities similar to other Class 0 protostars in the VANDAM Orion sample, is marginally resolved in the ALMA and VLA data, and is driving a strong outflow (Figs.~\ref{fig:fluxes} and \ref{fig:400outflow}). Its millimeter emission likely arises from a marginally resolved disk (Tobin et al. accepted). We interpret this protostar as a second stage HSC with a disk and outflow, similar to other Class 0 protostars in the VANDAM Orion sample. HOPS~400-A must have an unresolved temperature gradient and inner optically thick regions or disk. HOPS 400-A and -B are likely embedded in the same infalling, protostellar envelope. In the $Spitzer$, $Herschel$, and APEX data, they are not resolved and the resulting SED is a composite of the two \citep{furlan2016}.


HOPS 400-A has a bent morphology in the outflow as seen in Figure \ref{fig:400outflow}. Assuming that this bent nature is due to the motion of the protostar through the surrounding envelope. We estimate a velocity of 2.6 km s$^{-1}$ relative to its natal core. The value of the velocity adopts the average maximum velocity of the outflow from the red and blue lobe, half of the spatial distance in the y-direction, and the average x-direction distance of the blue and red lobe. It is unclear from this data whether the bent morphology originates from the proper motion of the binary or if it originates from the orbital motion of 400-A with respect to 400-B. The dynamical age of the outflow is $\sim$700 years.

Another possibility is that HOPS~400-B has an outflow that is hidden near the large optically thick continuum resulting in a superposition of two outflows. If this is the case, the outflows cannot be distinguished by the data presented in this paper and the outflow properties in Table \ref{tbl:outflow} may be a combination of two outflows instead of one. 

\subsection{HOPS~400-B}

The southern companion to HOPS~400-A extends almost 200~AU in diameter, at both 0.87~mm and 8~mm, with a peak to the south. The peak is more evident in the 8~mm band and is offset from the geometric center of the mass distribution; this is contrary to the nominal expectation of a protostar forming at the center of its envelope or disk. The brightness temperature profiles in Figure \ref{fig:temps} show a compact, central peak at 8~mm. This indicates the presence of a warm, compact density peak that is not apparent in the optically thick 0.87~mm emission. However, it is not clear if HOPS~400-B contains a HSC or not. HOPS~400-B is different than typical isothermal starless/pre-stellar cores in that the compression of gas in the opaque region is generating a significant amount of luminosity due to the high density of the gas. It is also possible, that HOPS~400-B contains a HSC and has launched an outflow that has not yet broken through the inner, dense envelope and therefore cannot be detected with the current ALMA data. Longer wavelength ALMA data may detect such an outflow in CO (J=1-0) due to the lower opacity at 2.7 mm.

\subsection{HOPS~401}

HOPS~401 is located within 15,000~AU of the irregular protostar HOPS~402, and like its neighbor, HOPS~401 has a very cold SED and lacks a detection at 24~$\mu$m \citep{stutz2013,furlan2016}. It is more compact than the irregular PBRs, spatially extended at 0.87~mm, barely resolved at 8~mm, and appears to be optically thick within the inner 29~AU at 0.87~mm (Table~\ref{tbl:pbrsproperties}). HOPS~401 either has a temperature gradient or an optical depth at 8~mm that approaches 1. The flux densities are similar to the  rest of the PBRs sample. Despite having 8~mm flux densities comparable to the typical PBRs, we include HOPS~401 in this analysis due to its cold SED, proximity to HOPS~402, and its hint of an irregular morphology.

There is high velocity $^{12}$CO emission near the source although it is not clear from our ALMA snapshot data if this is from an outflow driven by the protostar, an outflow from another protostar, or motions in the surrounding cloud (Fig. \ref{fig:401outflow}). The outflow lobes are not as distinct, are not exactly centered on the protostar, and have a bent morphology in the outflow similar to HOPS~400-A. If the bent nature is due to the motion of the protostar the estimated velocity of $\sim$7.5 km s$^{-1}$ relative to its natal core. The procedure to determine the velocity value is the same as for HOPS~400-A. The dynamical age of the outflow is $\sim$300 years.


Unlike the other four protostars with resolved optically thick zones, the free fall time of HOPS~401 is consistent with the Kelvin-Helmholtz timescale due to its lower inferred mass. Although its spatially resolved morphology appears irregular and not disk-like, HOPS~401 is not as extended at 8~mm as the other four sources. We did not include it in our primary analysis for this reason, yet it likely traces a very early stage of protostellar evolution and perhaps represents a pathway for hydrostatic core formation that does not produce a large optically thick zone. If this is the case, then the required duration for the four irregular protostars is higher than the value of $t_{stat}$ estimated in \S~\ref{sec:stat}.

\subsection{HOPS~402}

HOPS~402 has an asymmetric structure at 0.87~mm and 8~mm extending 150~AU at 0.87~mm. The 8~mm data has an elongated structure with a peak to the south and a triangular-shaped morphology coincident with the more extended 0.87~mm peak.  The brightness cut in Figure \ref{fig:temps} shows an extended and asymmetric structure with a compact 8~mm peak that is not visible at 0.87~mm. Here the peak 0.87~mm continuum is brighter, but the temperatures become comparable at the peak, again the signature of high optical depth at 0.87~mm.  

Similar to HOPS~400-B, it is not clear if HOPS~402 contains a HSC or not. There is no compact peak at either wavelength, but the compression of gas in the opaque region is generating a significant amount of luminosity due to the high density of the gas. Although the luminosities and $T_{bol}$ are the lowest of the PBR sample, it is possible HOPS~402 contains a HSC. This HSC may have launched an outflow that has not broken through the high opacity region and cannot be detected with the current ALMA data.

\subsection{HOPS~403}

HOPS~403 at 0.87~mm exhibits a roughly circular symmetry, extends over 150 AU in diameter, and has an elongated peak to the south-east. It shows a point-like peak at 8~mm that is coincident with the more elongated, lower contrast 0.87~mm peak. HOPS~403 is asymmetric in both wavelength cuts (Figure \ref{fig:temps}) with the 0.87~mm peak offset from the 8~mm peak. The 8~mm peak has an almost point-like shape but widens out past $\sim$40~AU. The fact that the 8~mm brightness temperature at this peak is almost double the 0.87~mm brightness temperature again indicates a high optical depth in the sub-mm with the 0.87~mm tracing cooler dust. The 0.87~mm emission is flatter outside of $\sim$50~AU, but is still within a factor of two to the 8~mm brightness temperature. The 8~mm peak is thought to be the source of the outflow, which would be more evolved than a FHSC as inferred from the outflow velocities, although we cannot trace the outflow directly to the source due to the high opacity at 0.87 mm. The dynamical age of the outflow is $\sim$300 years.


A second, fainter 8~mm peak not present in the 0.87~mm supports that this source may be forming a binary. In this case, the bright 8~mm peak may be located inside the optically thick region, or it maybe located behind the region. In the latter case, this orientation would be similar to HOPS~400 if observed from an orientation where HOPS~400-A is behind 400-B. Both would be forming in the center of the same infalling envelope, and the observed SEDs from $Spitzer$, $Herschel$ and APEX would be a composite SED for this entire system \citep{furlan2016} .  


\subsection{HOPS~404}

HOPS~404 has a box-like morphology at 8~mm and is a FHSC candidate. Unlike the other three irregular PBRs, the 0.87~mm continuum is more circularly symmetric and at both wavelengths, the continuum is the most compact of the four irregular PBRs. 


The outflow morphology of HOPS 404 also shows a bent, poorly collimated outflow perhaps demonstrating a rotation or precession of the outflow, the latter being the case if the outflow is more aligned in the plane of the sky (inclination~$>$ 60$^{\circ}$). This will reduce the projected outflow lengths and consequently increase the resulting dynamical lifetime of the outflow to $\sim$1400 years, which is larger than the HOPS~400-A and HOPS~403 dynamical lifetimes ($\sim$500 years). These lifetimes should be considered lower limits; the faster HOPS~400-A and HOPS~403 outflows may not have been fully mapped by ALMA. Thus, the HOPS~400-A and HOPS~403 HSCs can be older than the HOPS~404 FHSC.

\bibliography{pbrs.bib}
\end{document}